\documentclass[a4paper,12pt]{article}
\usepackage{amssymb,amsmath}
\usepackage{epsfig}

\usepackage{a4}
\usepackage{parskip}
\newcommand{\sect}[1]{\setcounter{equation}{0}\section{#1}}

\def\N{{\mathcal N}}
\def\L{{\mathcal L}}

\def\J{{\mathcal J}}
\def\O{{\mathcal O}}
\def\L{{\mathcal L}}
\def\A{{\mathcal A}}

\def\ds{\displaystyle}
\def\r{\rho}
\def\a{\alpha}
\def\b{\beta}
\def\d{\delta}

\def\s{\sigma}
\def\t{\tilde}

\def\m{\mu}
\def\n{\nu}

\def\f{\phi}
\def\vf{\varphi}
\def\e{\eta}
\def\ep{\varepsilon}

\def\l{\lambda}

\def\cosh{\mathrm{cosh}}
\def\arctanh{\mathrm{arctanh}}
\def\p{\partial}
\def\rb{\right}
\def\lb{\left}
\def\axs{AdS_5\times S^5}
\newcommand{\eq}[1]{\begin{equation} #1 \end{equation}}
\newcommand{\al}[1]{\begin{align} #1 \end{align}}
\newcommand{\ml}[1]{\begin{multline} #1 \end{multline}}

\begin{document}

\begin{titlepage}

\begin{flushright}{\small TUW--07--14}\end{flushright}

\vspace*{.5cm}
\begin{center}
{\Large \bf On the anatomy of  multi-spin \ \\ \vspace*{.3cm}
magnon and single spike string solutions}

\vspace*{1cm}
 H. Dimov${}^{\star}$ and R.C. Rashkov${}^{\dagger}$\footnote{e-mail:
rash@hep.itp.tuwien.ac.at; on leave from Dept of Physics, Sofia
University, Bulgaria.}

\ \\

${}^{\star}$ \textit{Department of Mathematics, University of
Chemical Technology and Metallurgy,\\ 1756 Sofia, Bulgaria}

\ \\

${}^{\dagger}$ \textit{Institute for Theoretical Physics, Vienna
University of Technology,\\
Wiedner Hauptstr. 8-10, 1040 Vienna, Austria}

\end{center}

\vspace*{.8cm}

\baselineskip=18pt

\bigskip
\bigskip
\bigskip
\bigskip

\begin{abstract}

We study rigid string solutions rotating in $AdS_5\times S^5$
background. For particular values of the parameters of the
solutions we find multispin solutions corresponding to giant
magnons and single spike strings. We present an analysis of the
dispersion relations in the case of three spin solutions
distributed only in $S^5$ and the case of one spin in $AdS_5$ and
two spins in $S^5$. The possible relation of these string
solutions to gauge theory operators and spin chains are briefly discussed.

\end{abstract}

\end{titlepage}

\sect{Introduction}

Currently String/gauge theory duality, the idea of which is rooted
in \cite{'tHooft:1973jz}, is one of the hottest topics. The
explicit form of this idea was provided when Maldacena conjectured
the AdS/CFT correspondence \cite{holography}. Since then this
subject became a major research area and many fascinating
discoveries were made in the last decade. It provides a powerful
tool for studies and future advances in the understanding of gauge
theories at strong coupling.

The best studied example is based on superstring theory on $\axs$
background where the corresponding gauge theory is $\N=4$
superconformal. Recent advances on both, the string and the gauge
theory sides of the correspondence have indicated that type IIB
string theory may be integrable. This is important for the
following reasons. One of the predictions of the correspondence is
the equivalence between the spectrum of free string theory on
$\axs$ and the spectrum of anomalous dimensions of gauge invariant
operators in the planar $\N=4$ Supersymmetric Yang-Mills (SYM)
theory. In order to check this conjecture we need the full
spectrum of free string theory on a curved space, such as $\axs$,
which is still an unsolved problem. In certain limits, \cite{GKP},
allowing reliability of the semiclassical results on both sides,
the techniques of integrable systems have become useful in
studying the AdS/CFT correspondence in detail. Assuming that these
theories are integrable, the dynamics should be encoded in an
appropriate scattering matrix $S$. This can be interpreted from
both sides of the correspondence as follows. On the string side,
in the strong-coupling limit the $S$ matrix can be interpreted as
describing the two-body scattering of elementary excitations on
the worldsheet. When their worldsheet momenta become large, these
excitations can be described as special types of solitonic
solutions, or giant magnons, and the interpolating region is
described by the dynamics of the so-called near-flat-space regime
\cite{Hofman:2006xt,Swanson:2006dec}. On the gauge theory side,
the action of the dilatation operator on single-trace operators is
the same as that of a Hamiltonian acting on the states of a
certain spin chain \cite{Minahan:2002ve}. This turns out to be of
great advantage because one can diagonalize the matrix of
anomalous dimensions by using the algebraic Bethe ansatz technique
(see \cite{fadd} for a nice review on the algebraic Bethe ansatz).
Then one can look for the S-matrix determined in an involving way
by the interable structures. The  dispersionless scattering
described in this approximation hopefully can be extended to the
regions where, due to the nature of this duality, validity of the
consideration in on one side rules out the validity on the other
side of the correspondence. However employing a certain spin chain
as a mediator between the two sides of the correspondence seems
very helpful. Indeed, in several papers the relation between
strings, $N=4$ SYM theory and spin chains was established, see for
instance
\cite{kruczenski},\cite{dim-rash},\cite{lopez},\cite{tseytlin1}
and references therein. This idea opened the way for a remarkable
interplay between spin chains, gauge theories, string
theory\footnote{For very nice reviews on the subject with a
complete list of references see
\cite{beisphd},\cite{tseytrev},\cite{Plefka:2005bk}} and
integrability (the integrability of classical strings on $\axs$
was proven in \cite{bpr}).

All these advantages lead to the idea that it would be useful to
look for string solutions governing various corners of the spin
chain spectrum. The most studied cases were spin waves in the
long-wave approximation corresponding to rotating and pulsating
strings in certain limits, see for instance the reviews
\cite{beisphd},\cite{tseytrev},\cite{Plefka:2005bk} and references
therein. Another interesting case are the low lying spin chain
states corresponding to the magnon excitations. One class of
string solutions already presented in a number of papers is the
string theory on pp-wave backgrounds. The latter, although
interesting and important, describes point-like strings which are
only part of the whole picture.

Let us briefly describe the state of the art. One of the important
explicit examples of string solutions corresponding to gauge
states of interest are the so called ``giant magnon'' solutions.
The relation between spin chain "magnon" states and specific
rotating semiclassical string states on $R\times S^2$ was
suggested by Hofman and Maldacena \cite{Hofman:2006xt}(for an early attempt see also \cite{samuel}). 
As we already pointed out, the investigation of various string solutions
which have clear interpretation in terms of gauge theory operators
is very important and this motivated the authors of
\cite{Dorey1},\cite{Dorey2},\cite{AFZ},\cite{MTT} to generalize
the results to magnon bound states. The latter are dual to strings
on $R\times S^3$ with two non-vanishing angular momenta. Certainly
it is of great interest to generalize further these investigations
to the cases of multispin magnon states. Such generalization was
given in \cite{MTT} where  different giant magnon states with two
and more spins were found, moreover the authors considered a
string solutions moving on $AdS_3\times S^1$ i.e. having spins in
both the AdS and the spherical part of the background (see also
\cite{Ryang:2006yq}). These classical string solutions were
further generalized to include dynamics on the whole $S^5$
\cite{volovich}, \cite{Kruczenski:2006pk}  and in fact a method to
construct classical string solutions describing superposition of
arbitrary scattering and bound states was found
\cite{Kalousios:2006xy}. The semiclassical quantization of the
giant magnon solution was performed in
\cite{Papathanasiou:2007gd}.

As we discussed above, one of the most important characteristics
of the solutions of this type are the dispersion relations. To be
more concrete, in the case of one spin giant magnon the latter is
found in \cite{Hofman:2006xt} to be:

\eq{ E-J=\ds\frac{\sqrt{\l}}{\pi}|\sin\ds\frac{p}{2}| } where $p$
is the magnon momentum which on the string side is interpreted as
a difference in the angle $\phi$ (see \cite{Hofman:2006xt} for
details). In the multispin cases the $E-J$ relation were studied
both on the string \cite{Dorey2},\cite{AFZ},\cite{MTT} and spin
chain sides \cite{Dorey1}. We will only note that a natural way to
extend this analysis is to find giant magnon solutions in
backgrounds which, via the AdS/CFT correspondence, are dual to
less supersymmetric gauge theories \cite{Leigh:1995ep}, for
instance, the gamma deformed $\axs$ background found by Lunin and
Maldacena \cite{Lunin:2005jy}, \cite{Frolov}. Investigations in
this direction, were pursued in
\cite{Chu:2006ae},\cite{Bobev:2006fg} (see
\cite{froits}-\cite{McLoughlin:2006cg} for interesting work on
semiclassical string dual to marginally deformed $\N=4$ SYM and
\cite{Swanson:2007dh} for a review).


A more general class of classical string solutions with spikes
maybe of interest for the AdS/CFT correspondence since they
incorporate the giant magnon solutions as a subclass. These spiky
strings were constructed in the $AdS_5$ subspace of $AdS_5\times
S^5$ \cite{Kruczenski:2004wg} and it was argued that they
correspond to single trace operators with a large number of
derivatives. The spiky strings were generalized to include
dynamics in the $S^5$ part of $AdS_5\times S^5$
\cite{Ryang:2005yd} and in \cite{McLoughlin:2006tz} closed strings
with "kinks" were considered\footnote{For other spiky string
solutions see \cite{Mosaffa:2007}}.

Recently a new analysis of the class of spiky string  appeared, in
\cite{Ishizeki:2007we,BR0706}, namely,  infinitely wound string
solutions with single spikes on $S^2$ and $S^3$ were found. It can
be shown that these solutions can be found in a certain limit of
the parameters of a general rotating rigid string. While the
interpretation of the giant magnon solutions from the field theory
point of view is as of higher twist operators, the single spike
solutions do not seem to be directly related to some gauge theory
operators. Nevertheless one can speculate on this issue suggesting an interpretation in terms of 
a spin chain Hubbard model \cite{Ishizeki:2007we}.
This means  that we are in the antiferromagnetic phase
\cite{Roiban:2006jt,Beisert:06-nl} of the corresponding spin chain, but the
relation to field theory operators is still unclear.

The single spike solutions possess interesting properties which
for instance modify the dispersion relations known from the giant
magnon case. For these spiky configurations the energy and ``the
deficit angle'' $\Delta\phi$ are infinite, but their difference is
finite
\eq{
E-T\Delta\phi=\frac{\sqrt{\l}}{\pi}\left(\ds\frac{\pi}{2}-\theta_1\right)
\notag }
where $T=\frac{\sqrt{\l}}{2\pi}$ is the string tension. The
angular momenta of the solution are finite and in the case of
single spike solution on $S^3$ their relation
\begin{equation}
J_1 = \sqrt{J_2^2 + \ds\frac{\lambda}{\pi^2} \cos^2\theta_1}
\end{equation}
resembles the dispersion relation of giant magnons on $S^3$. The
single spike solutions have relation to the sine-Gordon model and
this could be useful to study their scattering and eventually lead
to a better understanding of their properties. It is worth to
notice that one can generalize the single spike string solutions
to the case of the gamma deformed Lunin-Maldacena background
\cite{BR0706} following the analogy with the magnon case
\cite{Chu:2006ae, Bobev:2006fg}. The subject offers many directions for generalizations, for instance
in NS5-brane background \cite{Kluson:2007d5}. Classical membrane solutions with
spikes were investigated in \cite{bozhilov}.

It seems interesting to study the class of single spike solutions
in greater detail and better understand their properties. The
purpose of this paper is to present multispin single spike
solutions on the $\axs$ background which will generalize and
extend the result of \cite{Ishizeki:2007we,BR0706,Mosaffa:2007}.
Although there is no complete understanding and direct link to the
gauge theory operators, the correspondence suggests that they
should play important role at the high energy region. There might
be a connection to the Hubbard-like spin chain \cite{Roiban:2006jt,McLoughlin:2006cg,RSR}, but by now one can only
speculate on these issues.

The paper is organized as follows. In the next Section we concentrate on the
following questions. First of all we give 
a  concise review of the giant magnon string solutions. Next we present 
the scheme of reduction of the string sigma model to
Neumann-Rosochatius (NR) integrable system. We discuss in some
details  the conformal constraints which are intensively used
later. Next subsection has the purpose to demonstrate the derivation of
the discussed solutions by making use of NR integrable system. Note that 
although known, not all the solutions were obtained via this reduction. The idea is 
to treate all the cases using one approach - this of reduction to NR system.
In Section 3 we
find three spin single spike solutions in two cases. In the first
case we restrict the considerations to the case of string
configurations in $S^5$ part of the  space-time geometry. The
second type of solutions are strings extended in both, the
spherical and the $AdS$ part of the space-time geometry. In the
concluding section we discuss the results of our study.

\sect{One and two spin giant magnons and single spikes from strings on $\axs$}

In this section we present the  string solutions known as giant magnons and
single spikes with one and two spins. In the first subsection we review the idea 
of relating certain string solutions to magnon-like dispersion relations and
their relation to a class of gauge operators.  In the next subsection we present
in some details the reduction of
string sigma model to the so called Neumann-Rosochatius integrable system. The purpose of this
is to use an unified approach in treating the two classes of string solutions. 
Although the reviewed string solutions are known, not all of them are derived
using the reduction to NR integrable system In the next subsection we show how one can derive the reviewed 
string solution using the reduction to NR system. This can be considered as a worm-up for the derivations 
in the next section where we consider three spin cases.

\subsection{Review of one and two spin giant magnon solutions}

In this subsection we will present two spin string solutions
possessing spikes. These can be divided in two subclasses -
``giant magnon'' solutions and the so called ``single spike''
solutions. We start first with a short paragraph describing the
idea of giant magnons in string theory \cite{Hofman:2006xt} and
then present two spin solutions.

\paragraph{Giant magnons from strings on $\axs$} \ \\

First of all let us review the idea of Hofman and Maldacena
\cite{Hofman:2006xt} of obtaining ``giant magnon'' string
solutions. It comes with a simple analysis of the gauge theory
operators of specific type modeled by long spin chain composites.
Namely, let us look at a special class of gauge invariant
operators which are characterized by an infinite value of one of
the $SO(6)$ charges $J$ and have finite value of $E-J$. When $E-J$
vanishes the operator is composed of a chain of $J$ operators,
namely:

\eq{\O\sim Tr(ZZ..Z...Z)}

On this infinite "chain" we can consider a finite number of fields
$Y$ which can be interpreted as propagating excitations. The
corresponding operator takes the form:

\eq{\O\sim \ds\sum_{k}e^{ikp}(Z..ZZYZZ...Z)}

Where the summation is over all possible positions on the chain
where $Y$ can be inserted. This picture on the gauge side has a
description in terms of spin chains \cite{Minahan:2002ve}. The
operators $Z$ represent the ground state of the chain and the
$Y$'s correspond to magnon excitations.

Using supersymmetry arguments Beisert \cite{Beisert:2005tm} was
able to find the following dispersion relation for the magnon
excitations:

\eq{ E-J=\sqrt{1+\ds\frac{\l}{\pi^2}\sin^2\ds\frac{p}{2}}
\label{2.1}}

Here $p$ is the magnon momentum and the periodicity in $p$ is due
to the discreteness of the spin chain. The large 't Hooft coupling
limit of the above relation is:

\eq{E-J=\ds\frac{\sqrt{\l}}{\pi}\mid\sin\ds\frac{p}{2}\mid
\label{2.2}}

Hofman and Maldacena were able to reproduce the above relation
from string theory on a $R\times S^2$ subspace of $\axs$. They
considered a simple string with both ends rotating on the equator
of $S^2$ with constant separation between the ends
$\Delta\varphi$. The crucial step was to identify this angle
difference with the momentum of the spin chain magnon
$\Delta\varphi=p$. In order to be able to reproduce exactly the
magnon dispersion relation the following limits are considered:

\eq{J\rightarrow\infty \qquad E-J={\mathrm{fixed}} \qquad
\l={\mathrm{fixed}} \qquad p={\mathrm{fixed}} }

 We note that a closed folded string
configuration with one angular momentum as the one considered in
\cite{GKP} will correspond to a state which is a superposition of
two magnons with opposite momenta. The solution we are interested
in lives on the $R\times S^2$ background with the metric:

\eq{ds^2=-dt^2+d\theta^2+\sin^2\theta d\varphi^2\label{2.3} }

We look for a rotating string solution of the Polyakov action of
this model of the form:

\eq{ t=\tau \qquad \theta=\theta(y) \qquad
\varphi=t+\t{\varphi}(y)\label{2.4} }

Where we have defined the variable $y=cx-dt$. Here $x$ is a world
sheet spatial coordinate with infinite range and $c$ and $d$ are
constants. Using the above ansatz it is not hard to integrate the
equations of motion and get:

\eq{ \cos\theta=\ds\frac{\cos\theta_0}{\cosh y} \qquad
\tan\t{\varphi}=\ds\frac{\tanh y}{\tan\theta_0} \label{2.5} }

The constants $c$ and $d$ are related to the integration constant
$\theta_0$ by:

\eq{\ds\frac{1}{c}=\cos\theta_0 \qquad\qquad
d=\tan\theta_0\label{2.6} }

Now it is easy to see that $c^2-d^2=1$, this provides us with the
necessary condition that the group velocity $v=\frac{d}{c}$ is
less than one. We are interested in the conserved charges for our
system:

\eq{E=\ds\frac{\sqrt{\l}}{2\pi}\ds\int_{-\infty}^{\infty}dx \qquad
J=\ds\frac{\sqrt{\l}}{2\pi}\ds\int_{-\infty}^{\infty}dx\sin^2
\theta \label{2.7} }

Using the above relation we see that

\eq{E-J=\ds\frac{\sqrt{\l}}{\pi}\cos\theta_0 \label{2.8} }

The crucial part of the proposal of Hofman and Maldacena
\cite{Hofman:2006xt} is to make the identification
$\cos\theta_0=\sin(p/2)$. This leads to the conclusion that the
dispersion relations of the magnon excitation of the infinite spin
chain in the large 't Hooft coupling limit and the rotating string
with one large angular momentum are the same. Thus we have
outlined the idea in \cite{Hofman:2006xt} for finding the string
dual description of a giant magnon excitation in a certain limit.
This opens the possibility for generalizing this construction and
studying multi magnon bound states, which are argued to be dual
\cite{Dorey1} to rotating strings with three angular momenta. We
proceed with this analysis in the following sections.


\paragraph{Two spin giant magnons in $R\times S^3$} \ \\

Here we will present the two spin rotating string solution found
in \cite{AFZ} in a slightly different notation which will be more
convenient when we analyze the further generalizations. Another
advantage is that we will able to describe all the solutions of
this type in unified way, namely using appropriate limits of the
resulting NR reduction. The rotating string of interest is dual to
a bound state of $J_2$ magnons with charge $J_1$. We look for
classical string solution on a $R\times S^3$ subspace of $\axs$:

\eq{ ds^2=-dt^2+d\theta^2+\sin^2\theta d\phi_1^2+\cos^2\theta
d\phi_2^2\label{2.9} }

The Polyakov action for this system is:

\ml{ S_P=\ds\frac{\sqrt{\l}}{4\pi}\int d\tau d\s
[-(\p_{\tau}t)^2-(\p_{\tau}\theta)^2+(\p_{\s}\theta)^2
 +\sin^2\theta((\p_{\s}\phi_1)^2-(\p_{\tau}\phi_1)^2)+\\
\cos^2\theta((\p_{\s}\phi_2)^2-(\p_{\tau}\phi_2)^2)] \label{2.10}
}

It can be checked that the following ansatz is consistent with the
equations of motion following from the above action and the
Virasoro constraints:

\eq{\theta=\theta(y) \qquad \phi_1=t+g_1(y) \qquad \phi_2=\n
t+g_2(y) \label{2.11} } Here we have defined a new variable
$y=c\s-d\tau$ and we work in the conformal gauge $t=\tau$. It is
easy to derive the following equations for the unknown functions
$g_1(y)$,$g_2(y)$ and $\theta(y)$:

\eq{\begin{array}{l} \p_yg_1=\ds\frac{d}{\sin^2\theta}-d \qquad
\p_yg_2=-\n d\\\\
\p_y\theta=\cos\theta\ds\sqrt{1-\n^2
c^2-d^2\ds\frac{\cos^2\theta}{\sin^2\theta}}
\end{array}
\label{2.12} }

We can integrate the last equation and find $\cos\theta$ to be:

\eq{
\cos\theta=\sqrt{\ds\frac{1-\n^2c^2}{c^2-\n^2c^2}}\ds\frac{1}{\cosh(\sqrt{1-\n^2
c^2}y)} \label{2.13} }

It is useful to introduce the notation \eq{
\cos\theta_0=\sqrt{\ds\frac{1-\n^2c^2}{c^2-\n^2c^2}} \label{2.14}
} and following \cite{Hofman:2006xt}, we identify
$\sin\frac{p}{2}$ with $\cos\theta_0$, where $p$ is the momentum
of the magnon (interpreted as a geometrical angle in the string
picture). It is clear that once we know the solution $\theta(y)$
one can find explicitly the conserved charges of the rotating
string:

\eq{\begin{array}{l}
E=\ds\frac{\sqrt{\l}}{2\pi}\ds\int_{-\infty}^{\infty}d\s\\\\
J_1=\ds\frac{\sqrt{\l}}{2\pi}\ds\int_{-\infty}^{\infty}d\s
\sin^2\theta\left(1-d\p_yg_1\right)\\\\
J_2=\ds\frac{\sqrt{\l}}{2\pi}\ds\int_{-\infty}^{\infty}d\s
\cos^2\theta\left(\n-d\p_yg_2\right)
\end{array}\label{2.15}
}

After a little algebra we arrive at:

\eq{
E-J_1=\ds\frac{J_2}{\n}=\ds\frac{\sqrt{\l}}{\pi}\ds\frac{c\cos^2\theta_0}{\sqrt{1-\n^2c^2}}
\label{2.16}}

From here we extract the dispersion relation for the magnon bound
state found in \cite{Dorey1} and derived from the string sigma
model on $R\times S^3$ in \cite{AFZ},\cite{MTT}:

\eq{
E-J_1=\sqrt{J_2^2+\ds\frac{\l}{\pi^2}\sin^2\frac{p}{2}}\label{2.17}
}

The approach we used here was based on the original considerations
in \cite{Hofman:2006xt}.

\subsection{Reduction to Neumann-Rosochatius integrable system}

Motivated by the considerations made in the case of magnon string
configurations \cite{Kruczenski:2006pk}, we find it useful to
apply the generalized Neumann-Rosochatius(NR) ansatz
\cite{arts-new-int} \eq{ X_a=x_a(\xi)e^{i\omega_a\tau}, \quad
\xi=\a\sigma+\b\tau, \quad x_a(\xi+2\pi\a)=x_a(\xi), \quad a=1,2,3
\label{period-x} } for the spherical part of the geometry and \eq{
Y_l=\m_l(\xi)e^{iv_l\tau}, \quad \m_l(\xi+2\pi\a)=\m_l(\xi), \quad
l=0,1,2 \label{period-y} } for the $AdS_5$ piece. Let us restrict
our attention here to the spherical part, we will discuss the
other part in subsequent sections. The embedding we choose
supplies the theory with two types of constraints - those coming
from the embedding, namely $X_aX_b\eta^{ab}=1$, and the conformal
constraints. The latter are very important when we are dealing
with the Polyakov string action, so let us briefly consider them
here.

\paragraph{Conformal constraints in NR} \ \\

In general the Virasoro constraints can be written as \al{
& \sum\limits_a\lb[|\p_\tau X_a|^2+|\p_\sigma X_a|^2\rb]=\kappa^2 \label{vir1} \\
& \sum\limits_a\lb[ \p_\tau X_a\p_\sigma\bar X_a+\p_\tau\bar
X_a\p_\sigma X_a\rb]=0 \label{vir2} } where we used that
$t=\kappa\tau$ and the rest of the AdS part is turned off.
 Using that
 \al{ \p_\tau X_a=(\b x_a'+i\omega_a x_a)e^{i\omega_a\tau}, \quad \p_\sigma X_a=\a
x_a'e^{i\omega_a\tau}; \quad {}'\equiv \ds\frac{d}{d\xi} \notag }
one finds for (\ref{vir1}, \ref{vir2}) \al{
& \sum\limits_a\lb[(\b x_a'+i\omega_a x_a)(\b\bar x_a'-i\omega_a\bar x_a)+
\a^2 x_a'\bar x_a'\rb]=\kappa^2 \label{vir1-a} \\
& \sum\limits_a\lb[(\b x_a'+i\omega_a x_a)\a\bar x_a'+ \a
x_a'(\b\bar x_a'-i\omega_a\bar x_a)\rb]=0 \label{vir2-a} } Our
starting point will be (\ref{vir2-a}) which can be rewritten as
\eq{ \sum\limits_a\lb[2\b x_a'\bar x_a'+i\omega_a(x_a\bar x_a'-
x_a'\bar x_a)\rb]=0 \label{vir2-b} } For notational simplicity we
define \eq{ \Xi_a=i(x_a\bar x_a'- x_a'\bar x_a) } With the help of
(\ref{vir2-b}) one can eliminate from (\ref{vir1-a}) the
combination $\Xi_a$. The result is \eq{ (\a^2-\b^2)\sum\limits_a
x_a'\bar x_a'+\sum\limits_a\omega_a^2 x_a\bar x_a=\kappa^2
\label{vir1-b} } To obtain purely algebraic relation one can
eliminating $x_a'\bar x_a'$ from (\ref{vir1-a}) to find \eq{
-\frac{(\a^2-\b^2)}{2\b}\sum\limits_a \omega_a\Xi_a+\sum\limits_a
\omega_a^2 x_a\bar x_a=\kappa^2 \label{vir2-c} } After all these
manipulations let us write down the final expressions we will use
\al{
&(\a^2-\b^2)\sum\limits_a x_a'\bar x_a'+\sum\limits_a\omega_a^2 x_a\bar x_a=\kappa^2 \label{vir1-fin} \\
&-\frac{(\a^2-\b^2)}{2\b}\sum\limits_a \omega_a\Xi_a+\sum\limits_a
\omega_a^2 x_a\bar x_a=\kappa^2 \label{vir2-fin} }

\paragraph{The Lagrangian} \ \\

The consistent reduction to the Neumann-Rosochatius interable
system requires further specification of the parametrization
(\ref{period-x}). The parametrization we need is defined as \eq{
x_a(\xi)=r_a(\xi)e^{i\phi_a(\xi)} , \quad
|x_a'|^2={r_a'}^2+r_a^2{\phi_a'}^2, \quad \Xi_a=2r^2_a\phi_a'
\label{parametr} } In these variables the Lagrangian takes the
following form \ml{
\L=\sum\limits_a\lb[(\a^2-\b^2){r_a'}^2+(\a^2-\b^2)r_a^2\lb(\phi_a'-\frac{\b\omega_a}{\a^2-\b^2}\rb)^2-
\frac{\a^2}{\a^2-\b^2}\omega_a^2r_a^2\rb]\\
+\Lambda\lb(\sum\limits_a r_a^2-1\rb) \label{lagr-1} } It is easy
to eliminate the angular degrees of freedom $\phi_a$. First we can
integrate the equations of motion for $\phi_a$ to get \eq{
\phi_a'=\frac{1}{\a^2-\b^2}\lb(\frac{C_a}{r_a^2}+\b\omega_a\rb)
\label{eq-phi} } where $C_a$ are constants of motion. On other
hand for $r_a$ we get \eq{
(\a^2-\b^2)r_a''-\frac{C_a^2}{\a^2-\b^2}\frac{1}{r_a^3}+\frac{\a^2}{\a^2-\b^2}\omega_a^2r_a-\Lambda
r_a=0 \label{eq-r} } where we used the equations for $\phi_a$
(\ref{eq-phi}). The net result of this considerations is that we
reduced the essential degrees of freedom to $r_a$.

The next step is to obtain the effective theory for $r_a$. The
equations of motion for $r_a$ can be
 obtained from the following effective Lagrangian
\eq{
\L=\sum\limits_a\lb[(\a^2-\b^2){r_a'}^2-\frac{C_a^2}{\a^2-\b^2}\frac{1}{r_a^2}-\frac{\a^2}{\a^2-\b^2}\omega_a^2r_a^2\rb]
+\Lambda\lb(\sum\limits_a r_a^2-1\rb) \label{lagr-nr} } which is
nothing but the well known integrable Neumann-Rosochatius system
\cite{arts-new-int}.
 One can easily find the corresponding on-shell Hamiltonian which is one of the classical constants of motion
\eq{  H=
\sum\limits_a\lb[(\a^2-\b^2){r_a'}^2+\frac{C_a^2}{\a^2-\b^2}\frac{1}{r_a^2}
+\frac{\a^2}{\a^2-\b^2}\omega_a^2r_a^2\rb]
\label{hamil-1}
}
The considerations from now on will be essentially based on the above reduction.

\paragraph{The conformal constraints - second round} \ \\

As we argued above, the reduction to the string theory with
desired properties requires certain ansatz for the string
embedding coordinates. Let us examine how the Virasoro constraints
look like using this ansatz, this will play an essential role in
what follows.

Using the definition $\Xi_a=2r_a^2\phi_a'$ and the equations of
motion for $\phi_a$ (\ref{eq-phi}), the second constraint
(\ref{vir2-fin}) can be written in the form \al{
& \Xi_a=\frac{2}{\a^2-\b^2}\lb[C_a+\b\omega_ar_a^2\rb], \notag \\
&\text{after substitution in (\ref{vir2-fin})}\,\, \Longrightarrow
\sum\limits_a\omega_aC_a+\b\kappa^2=0. \label{vir2-nr} } In more
details, using that \eq{
|x_a'|^2=r_a'^2+\frac{1}{(\a^2-\b^2)^2}\lb[\frac{C_a^2}{r_a^2}+\b^2\omega_a^2r_a^2+2\b
C_a\omega_a\rb] }
 one can rewrite the first Virasoro constraint (\ref{vir1-fin}) as
\eq{
\sum\limits_a\lb[(\a^2-\b^2){r_a'}^2+\frac{C_a^2}{\a^2-\b^2}\frac{1}{r_a^2}
+\frac{\a^2}{\a^2-\b^2}\omega_a^2r_a^2+\frac{2\b
C_a\omega_a}{\a^2-\b^2}\rb]=\kappa^2 \label{vir1-fin-1}. }
 Comparing with (\ref{hamil-1}) we observe that the first three terms are
exactly the on-shell Hamiltonian, i.e. \eq{
H+\sum\limits_a\frac{2\b C_a\omega_a}{\a^2-\b^2}=\kappa^2 }
 But according to the second constraint (\ref{vir2-nr}) we can rewrite
the above equation as \eq{
H=\kappa^2+\frac{2\b^2}{\a^2-\b^2}\kappa^2=\frac{\a^2+\b^2}{\a^2-\b^2}\kappa^2
} One can write then the final form of the Virasoro constraints as
\al{
& H=\frac{\a^2+\b^2}{\a^2-\b^2}\kappa^2 \label{vir1-fin-nr} \\
& \sum\limits_a\omega_aC_a+\b\kappa^2=0 \label{vir2-fin-nr} }

We should also impose periodicity conditions on $r_a$ and
$\phi_a$:
 \eq{ r_a(\xi+2\pi\a)=r_a(\xi), \quad
\phi_a(\xi+2\pi\a)=\phi_a(\xi)+2\pi n_a, \,\, n_a -
\text{integer.} }

\subsection{One and two spin cases from NR system}

As we discussed above, one can formulate the problem in
terms of NR integrable system. In this subsection should be considered as a preparation 
for the studies in the next section.

Let us consider directly the two spin case (the single spin one can be obtained as a
simple limit \cite{Kruczenski:2006pk}).
Since we are dealing with the two
spin case, the considerations are restricted to $S^3\subset S^5$,
i.e. to only one independent radial variable, say $r_1$
($r_1^2+r_2^2=1$). One can find the first order equation for $r_1$
either from the equations of motion(\ref{eq-r}) (integrating them
once) or from the Virasoro constraints. The result
is\footnote{Here we set $\a=1$, which does not affect the
considerations}
\ml{
(1-\b^2)^2{r_1'}^2=\frac{1}{r_1^2}[((1+\b^2)\kappa^2-
\omega_2^2)(1-r_1^2)r_1^2-C_1^2\\
+(C_1^2-C_2^2)r_1^2-(\omega_1^2-\omega_2^2)(1-r_1^2)r_1^4]
\label{eq-r-2sp}
}
The right hand side determines the turning
points ${r_1'}^2=0$ and they are three. In order the string to
extends to the equator of the sphere, one must choose $r_1=1$.
This choice fixes $C_2=0$ (which can be deduced also requiring
regularity at $r_1=1$ for the equations of motion). To find a
solution of the type we are looking for, $r_1=1$ has to be double
zero of the right hand side of (\ref{eq-r-2sp}). The latter
conditions leads to the following constraints
\eq{
(1+\b^2)\kappa^2=\omega_1^2+C_1^2, \quad \omega_1 C_1+\omega_2 C_2
+\b\kappa^2=0
\label{eq-param-nr}
}
which can be obtained either
by substituting $r_1=1$ in the right hand side of (\ref{eq-r-2sp})
or from the Virasoro constraints. The correct choice for the
parameters solving the above equation and giving magnon type
string solutions is\footnote{See the Appendix for notations and
details.}
\eq{ \kappa=\omega_1, \quad \a=1, \quad
\b=-\frac{C_1}{\omega_1}
\label{param-2sp}
}
With this choice the equation for $r_1$ becomes
\eq{
(1-\b^2)^2{r_1'}^2=\frac{(1-r_1^2)^2}{r_1^2}\Delta\omega^2(r_1^2-\bar
r_1^2)
\label{star1} }
Leaving the details for the Appendix, the
final result for the dispersion relation is
\eq{
E-J_1=\sqrt{J_2^2+\ds\frac{\l}{\pi^2}\sin^2\frac{p}{2}}
}
which,
of course, is the same as (\ref{2.17}).

\paragraph{The solution in $SL(2)$ sector} \ \\

It is straightforward to apply the same ideology and obtain
analogous solutions in the AdS part of the geometry (including
also one angle from the spherical part, $\tilde\f$). Let us demonstrate how one can
find the corresponding result from \cite{Kruczenski:2006pk} using the above approach. 
In terms of NR system, to
obtain the solutions in question we make the ansatz
\al{
& \m_0=\m_0(\sigma), \quad \tilde\vf=v\tau +\vf(\sigma), \quad t=\kappa\tau\notag \\
& \m_1=\m_1(\sigma), \quad \tilde\f=\omega\tau +\f(\sigma), \quad
\tilde\f\in S^5
}
Then the Virasoro constraints become
\al{
& -\m_0'^2-\kappa^2\m_0^2+\m_1'^2+\m_1^2\vf'^2+ v^2\m_1^2+\phi'^2+
\omega^2=0 \label{slvir1} \\
& v\m_1^2\vf'+\omega\f'=0
\label{slvir2}
}
From here one can
easily find
\eq{ \vf'=-\frac{\omega}{v\m_1^2}\f' \label{vf-f} }
as expected.

Now one can solve for $\m_0$ or $\m_1$ ($\m_0^2-\m_1^2=1$) to
obtain the result for $SL(2)$ sector. To reproduce the results of
\cite{MTT,Kruczenski:2006pk} we fix the following values for the
parameters
\eq{ \omega=v=\kappa=1 }
 For this fixing the first Virasoro constraint becomes
\eq{
-\m_0'^2-\m_0^2+\frac{\m_0^2\m_0'^2}{\m_0^2-1}+\frac{\m_0^2\f'^2}{\m_0^2-1}+\m_0^2=0
}
where we used the constraint $\m_0^2-\m_1^2=1$. Using
(\ref{vf-f}) again, one can  find
\eq{
\m_0=\frac{\sin\vf_0}{\sin\vf}, \quad -\vf_0<\vf<\vf_0
}
which is the same as eq(C.6) of \cite{Kruczenski:2006pk}. The quantum
numbers of interest are given by
\al{
& E-J=\frac{\sqrt{\l}}{2\pi(1-\omega^2)}\int\limits^{\vf_0}_{-\vf_0}\frac{\sin\vf_0}{\sin^2\vf} \notag \\
& S=\omega(E-J).
\label{sl-disp}
}
Note that in order to find the correct dispersion relations one must regularize the AdS spin $S$
(see \cite{MTT} for details). The final result is
\eq{
(E-J)_{reg}=-\sqrt{|S_{reg}|^2+\frac{\l}{\pi^2}\sin^2\frac{p}{2}}
}

\paragraph{Two-spin case - spiky solutions} \ \\

Let us turn to the two spin solutions developing a single spike.
This configuration can be realized in terms of the NR integrable
system with specific choice of the parameters. The solutions we
are looking for are characterized by large quantum numbers,
especially large energy. The careful analysis shows (see for
instance  \cite{Kruczenski:2006pk,BR0706}) that in order to have
such solutions one has to choose the parameters in a specific way.

The ``spiky'' choice for the parameters, namely the choice giving
solutions with a single spike but infinitely wound around the
equator, is slightly different from the case of giant magnons. In
fact the constraints on the parameters are the same,
(\ref{eq-param-nr}), but instead of the choice (\ref{param-2sp}),
now we choose the other solution to the constraint
(\ref{eq-param-nr})
\eq{ \kappa=C_1, \quad
\beta=-\frac{\omega_1C_1}{\kappa^2}=-\frac{\omega_1}{C_1}
\label{param-ssp}
}
The equation for the variable $r_1$ is the
same (\ref{star1}), but the parameters are different
\eq{
\frac{d\,u}{d\xi}=u'=\frac{\sqrt{\Delta\omega^2}}{1-\beta^2}(1-u)\sqrt{u-\bar
u}.
}
Above we use the following notations
\al{ &
u=r_1^2=\sin^2\theta, \quad \bar
u=\frac{C_1^2}{\sqrt{\Delta\omega^2}}, \quad
 \Delta\omega^2 = \omega_1^2-\omega_2^2,  \notag \\
& d\xi=\frac{du}{u'}=\frac{(1-\beta^2)\,du}{2\sqrt{\Delta\omega^2}(1-u)
\sqrt{u-\bar u}}= \frac{(C_1^2-\omega_1^2)\,d
u}{2C_1^2\sqrt{\Delta\omega^2}(1-u)\sqrt{u-\bar u}}
}

The conserved quantities are (see equations (3.7)-(3.10)
\al{
& \Delta\phi_1=\frac{C_1}{(1-\beta^2)}\int\frac{d\xi}{u} +
\frac{\beta\omega_1}{(1-\beta^2)}\int d\xi \notag \\
& E=\kappa T\int d\xi \notag \\
& J_1=\frac{C_1\beta}{(1-\beta^2)}\int d\xi + \frac{\omega_1}{(1-\beta^2)}T\int u\,d\xi \\
& J_2=\frac{\omega_2}{(1-\beta^2)}T\int(1-u)d\xi. \notag
}

To find finite results (which is so for $E-J_1$ in magnon case) we
consider
\eq{
E-T\Delta\phi_1=\frac{2C_1T}{\sqrt{\Delta\omega^2}}\arccos\sqrt{\bar
u}=\frac{\sqrt{\lambda}}{\pi}\bar\theta
}
where
\eq{
\bar\theta=\frac{\pi}{2}-\theta_0
}
For the spins we get
\eq{
J_1=\frac{2T\omega_1}{\sqrt{\Delta\omega^2}}\cos\theta_0=\frac{2T\omega_1}
{\sqrt{\Delta\omega^2}}\sin\bar\theta
}
Analogously
\eq{
J_2=-\frac{2T\omega_2}{\sqrt{\Delta\omega^2}}\cos\theta_0=-\frac{2T\omega_2}{\sqrt{\Delta\omega^2}}
\sin\bar\theta
}
Defining $\sin\gamma$ as
\eq{
\sin\gamma=\frac{\omega_2}{\omega_1},
\sin\theta_0=\frac{C_1}{\sqrt{\Delta\omega^2}}
}
we find
\al{
& J_1=2T\frac{1}{\sin\gamma}\sin\bar\theta \notag \\
& J_2=-2T\frac{\sin\gamma}{\cos\gamma}\sin\bar\theta
}
These give the result for dispersion relations as in
\cite{Ishizeki:2007we,BR0706}
\al{
& E-T\Delta\phi_1=\frac{\sqrt{\lambda}}{\pi}\bar\theta \\
& J_1^2=J_2^2+\frac{\lambda}{\pi^2}\sin^2\bar\theta
}
We see that setting $J_2=0$ reproduce the one spin case.


\sect{Three-spin case}

As we discussed in the Introduction, the complete string solution in the full
$AdS_5\times S^5$ geometry is
not yet available and the thorough analysis of string solutions of the type
we are interesting in is important for description of different corners of the
correspondence. Such studies of the string solutions can also
serve as strong clue towards obtaining the complete solutions in
question.

Having the analysis above it is reasonable to ask whether one can
use the NR system to describe all these solutions and find the
corresponding dispersion relations. In the case of giant magnons
the answer is positive and given in \cite{Kruczenski:2006pk}. For
single spike it is an open question. The two types of solutions we
discussed belong to different corners of the energy scale. Then an
interesting question is is it possible to combine the magnon and
single spike solution and how such combination would look like?

In this section first we will review the three spin giant magnon
solution and will find new single spike solutions with three
spins. We will also consider the three spin solution which is a
combination of a two spin single spike on $S^5$ and a giant magnon
on $AdS_5$.

\subsection{The solution with three angular momenta in $S^5$}

In this subsection we give a brief review of the giant magnon
solution with three non-zero angular momenta obtained in
\cite{Kruczenski:2006pk} and extend it to the case of ``single
spike'' string solutions. The derivation uses the reduction to the
Neumann-Rosochatius integrable system. With certain choice of the
parameters one can find giant magnon solutions. Here we will
obtain the choice of the parameters, solutions of the imposed
constraints, so that the resulting solutions are of ``single
spike'' type.

The standard approach in solving the NR system is the change the
variables passing from three constrained radial variables $r_a$ to
two unconstrained ones, $\zeta_\pm$ (we put all
$\omega_a\neq\,0$). This is achieved by introducing ellipsoidal
coordinates through the relations:
\begin{equation}
\sum\limits_{a=1}^3\,\frac{r_a^2}{\zeta-\omega_a^2}=
\frac{(\zeta-\zeta_+)(\zeta-\zeta_-)}{\prod\limits_{a=1}^3\,(\zeta-\omega_a^2)}\,.\label{3s1}
\end{equation}
The two independent variables $\zeta_\pm$ are defined as the roots
of the quadratic equation obtained by taking the common
denominator on the left hand side and setting the numerator equal
to zero. The two roots are such that
$\omega_3^2<\zeta_-<\omega_2^2<\zeta_+<\omega_1^2\,$ and one can
check that they satisfy the relations:
\begin{equation}
\zeta_++\zeta_-=\sum\limits_a\omega_a^2-\sum\limits_a\omega_a^2r_a^2,\qquad
\zeta_+\zeta_-=\lb(\prod\limits_a\omega_a^2\rb).\sum\limits_b\frac{r_b^2}{\omega_b^2}.\label{3s2}
\end{equation}
The inverse  transformation can be easily obtained and it reads
\begin{equation}
r_a^2=\frac{(\zeta_+-\omega_a^2)(\zeta_--\omega_a^2)}{\prod\limits_{b\neq{a}}(\omega_a^2-
\omega_b^2)}, \quad a=1,2,3. \label{3s3}
\end{equation}
Having the correct coordinates respecting the symmetries of the
system, one can compute in a straightforward way the Hamiltonian
of the NR system in terms of $\zeta_{\pm}$:
\begin{equation}
H_\zeta=\frac{1}{(\alpha^2-\beta^2)(\zeta_+-\zeta_-)}
\left[\tilde{H}(p_-,\zeta_-)-\tilde{H}(p_+,\zeta_+)\right],\label{3s4}
\end{equation}
where
\begin{equation}
\tilde{H}(p,\zeta)=\prod\limits_{a}(\zeta-\omega_a^2)p^2+\sum\limits_{a}
C_a^2\frac{\prod\limits_{b\neq{a}}(\omega_a^2-\omega_b^2)}{\zeta-\omega_a^2}+
\alpha^2\sum\limits_{a}\omega_a^2\zeta-\alpha^2\zeta^2\,.
\label{3s5}
\end{equation}
The standard way to study this system is to use the
Hamilton-Jacobi method which allows separation of variables. This
approach requires finding a function $W(\zeta_+,\zeta_-)$ such
that\footnote{Here we follow the notation of
\cite{Kruczenski:2006pk}. }
\begin{equation}
H_{\zeta}\left(p_{\pm}=\frac{\partial{W}}{\partial{\zeta_{\pm}}},\,\zeta_{\pm}\right)=\,E\,.\label{3s6}
\end{equation}
If a solution of the form $W=W_+(\zeta_+)+W_-(\zeta_-)$ exist we
say that the variables separate and the system is integrable in
these coordinates\footnote{It is known that the
Neumann-Rosochatius system is integrable and therefore if the
choice of the coordinates is correct the separation should be
straightforward.}. Indeed, if we make this ansatz one can obtain
that these two functions satisfy the same equation, i.e.
 $W_\pm$ are the same function obtained from integrating the equation
\begin{equation}
\left(\frac{\partial{W}}{\partial{\zeta}}\right)^2=
\frac{V-\sum\limits_a{\prod\limits_{b\neq{a}}}(\omega_a^2-\omega_b^2)\frac{C_a^2}{\zeta-\omega_a^2}+
\left[\kappa^2(\alpha^2+\beta^2)-\alpha^2\sum\limits_{a}\omega_a^2\right]\zeta+\alpha^2\zeta^2}
{\prod\limits_{a}^{\,}(\zeta-\omega_a^2)},\label{3s7}
\end{equation}
In the above equation $V$ is a constant of motion and we have used
the expression for the hamiltonian \eqref{vir1-fin-nr} which gives
$E=\frac{\alpha^2+\beta^2}{\alpha^2-\beta^2}\,\kappa^2$. The
solution of the Hamilton-Jacobi equation is then
\begin{equation}
W(\zeta_\pm,V,E)=W(\zeta_+,V,E)+W(\zeta_-,V,E).\label{3s8}
\end{equation}
The equations of motion can be obtained using the standard
procedure
\begin{gather}
\frac{\partial{W(\zeta_+,V,E)}}{\partial{V}}+\frac{\partial{W(\zeta_-,V,E)}}{\partial{V}}=U,\label{3s9}\\
\frac{\partial{W(\zeta_+,V,E)}}{\partial{E}}+\frac{\partial{W(\zeta_-,V,E)}}{\partial{E}}=\xi,\label{3s10}
\end{gather}
where we introduced  a new constant $U$ and we see that $\xi$
plays the role of ``time'' variable. Taking derivatives of $W$ we
find
\begin{align}
&\int^{\zeta_+}\,\frac{d\zeta}{\sqrt{P_5(\zeta)}}+\int^{\zeta_-}\,\frac{d\zeta}{\sqrt{P_5(\zeta)}}=
2U,\label{3s11}\\
&\int^{\zeta_+}\,\frac{\zeta\,d\zeta}{\sqrt{P_5(\zeta)}}+\int^{\zeta_-}\,\frac{\zeta\,d\zeta}
{\sqrt{P_5(\zeta)}}= -\frac{2\xi}{\alpha^2-\beta^2},\label{3s12}.
\end{align}
Here the polynomial $P_5(\zeta)$ is the well known polynomial  of
degree five appearing in the $O(6)$ NR integrable system
\cite{arts-new-int}
\begin{multline}
P_5(\zeta)=\prod\limits_{a}(\zeta-\omega_a^2)\left\{V-\sum\limits_a{\prod\limits_{b\neq{a}}}
(\omega_a^2-\omega_b^2)
\frac{C_a^2}{\zeta-\omega_a^2}-\right.\\
\left.-\left[\kappa^2(\alpha^2+\beta^2)-\alpha^2\sum\limits_{a}\omega_a^2\right]
\zeta-\alpha^2\zeta^2\right\}. \label{3s13}
\end{multline}
The defined integrable system has many solutions which can be
obtained using various methods, for example dressing method,
B\"{a}cklund transformations etc.. Here we are interested however
in particular string solutions. In order to have reliable
semiclassical results and to be able to apply the AdS/CFT
correspondence we are looking for solutions describing strings
with large quantum numbers, say one infinite momentum or
"infinitely long" string (a single spike sting). For this to
happen the variables $\zeta_\pm$ must reach its extremal values
$\omega_{2,3}^2$. To satisfy this condition we are forced to
choose $V$ and the energy $E$ (or equivalently $\kappa$) in such a
way that $P_5(\zeta)$ has a double zero at
$\zeta_\pm=\omega_{2,3}^2$. For this to happen we are forced to
choose:
\begin{equation}
C_2=0,\,C_3=0,\,\kappa^2(1+\beta^2)=\omega_1^2+C_1^2,\,
V=-\omega_2^2\omega_3^2-C_1^2(\omega_1^2-\omega_2^2-\omega_3^2).\label{3s14}
\end{equation}
As in the previous section we put $\alpha=1$ without lost of
generality. This greatly reduce the complexity of the equations we
have to solve, namely
\begin{align}
&\int_{\bar{\zeta}}^{\zeta_+}\,\frac{d\zeta}{(\zeta-\omega_2^2)(\zeta-\omega_3^2)
\sqrt{\bar{\zeta}-\zeta}}+
\int_{\bar{\zeta}_-}^{\zeta_+}\,\frac{d\zeta}{(\zeta-\omega_2^2)(\zeta-\omega_3^2)
\sqrt{\bar{\zeta}-\zeta}}=0,
\label{3s15}\\
&\int_{\bar{\zeta}}^{\zeta_+}\,\frac{\zeta\,d\zeta}{(\zeta-\omega_2^2)(\zeta-\omega_3^2)
\sqrt{\bar{\zeta}-\zeta}}+
\int_{\bar{\zeta}_-}^{\zeta_+}\,\frac{\zeta\,d\zeta}{(\zeta-\omega_2^2)(\zeta-\omega_3^2)
\sqrt{\bar{\zeta}-\zeta}}= -\frac{2\xi}{1-\beta^2}.\label{3s16}
\end{align}
A few remarks are in order. We assume that the maximum value of
$\zeta_+$ is $\bar{\zeta}=\omega_1^2-C_1^2$ where its range is
$\omega_2^2<\bar{\zeta}<\omega_1^2$. In addition we assume that at
such point $\zeta_-$ has an arbitrary value
$\bar{\zeta}_-\,\,(\omega_3^2<\bar{\zeta}_-<\omega_2^2)$ which
means that we are free to choose the value of the constant $U$.
From now on we set $U=0$. Applying the above settings and using
simple manipulations we can rewrite the equations (\ref{3s15}) and
(\ref{3s16}) in a simplified form
\begin{align}
&\int_{\bar{\zeta}}^{\zeta_+}\,\frac{d\zeta}{(\zeta-\omega_3^2)\sqrt{\bar{\zeta}-\zeta}}+
\int_{\bar{\zeta}_-}^{\zeta_+}\,\frac{d\zeta}{(\zeta-\omega_3^2)\sqrt{\bar{\zeta}-\zeta}}=
-\frac{2\xi}{1-\beta^2},\label{3s17}\\
&\int_{\bar{\zeta}}^{\zeta_+}\,\frac{\zeta\,d\zeta}{(\zeta-\omega_2^2)\sqrt{\bar{\zeta}-\zeta}}+
\int_{\bar{\zeta}_-}^{\zeta_+}\,\frac{\zeta\,d\zeta}{(\zeta-\omega_2^2)\sqrt{\bar{\zeta}-\zeta}}=
-\frac{2\xi}{1-\beta^2}. \label{3s18}
\end{align}
One can integrate these equations using elementary methods. The
results are
\begin{align}
&\int_{\bar{\zeta}}^{\zeta_+}\,\frac{d\zeta}{(\zeta-\omega_2^2)\sqrt{\bar{\zeta}-\zeta}}=
\frac{2}{\sqrt{\bar{\zeta}-\omega_2^2}}
\arctanh\frac{\sqrt{\bar{\zeta}-\zeta_+}}{\sqrt{\bar{\zeta}-\omega_2^2}},\label{3s19}\\
&\int_{\bar{\zeta}_-}^{\zeta_-}\,\frac{d\zeta}{(\zeta-\omega_2^2)\sqrt{\bar{\zeta}-\zeta}}=
\frac{2}{\sqrt{\bar{\zeta}-\omega_2^2}}\left[
\arctanh\frac{\sqrt{\bar{\zeta}-\omega_2^2}}{\sqrt{\bar{\zeta}-\zeta_-}}
-\arctanh\frac{\sqrt{\bar{\zeta}-\omega_2^2}}{\sqrt{\bar{\zeta}-\bar{\zeta}_-}}\right].
\label{3s20}
\end{align}
Now our problem reduces to an algebraic one, by using the above
results we find
\begin{gather}
\frac{s_+s_-+s_2^2}{s_++s_-}=s_2A_2(\xi),\label{3s21}\\
\frac{s_+s_-+s_3^2}{s_++s_-}=s_3A_3(\xi),\label{3s22}
\end{gather}
To simplify further the notations we follow
\cite{Kruczenski:2006pk} and introduce the quantities
\begin{gather}
s_1=\sqrt{\omega_1^2-\bar{\zeta}},\,\,s_{2,3}=\sqrt{\bar{\zeta}-\omega_{2,3}^2},\,\,s_{\pm}=
\sqrt{\bar{\zeta}-\zeta_{\pm}},\label{3s23}\\
A_2(\xi)=\tanh\left(-\frac{s_2\xi}{1-\beta^2}+C_2\right),\,\,
A_3(\xi)=\coth\left(-\frac{s_3\xi}{1-\beta^2}+C_3\right).\label{3s24}.
\end{gather}
where $C_i$ stands for
\begin{equation}
\tanh{C_2}=\frac{s_2}{\sqrt{\bar{\zeta}-\bar{\zeta}_-}},\qquad
\tanh{C_3}=\frac{\sqrt{\bar{\zeta}-\bar{\zeta}_-}}{s_3},\label{3s25}
\end{equation}
and $\xi$ is assumed to take values in the range
$(-\infty,\infty)$. Since the quantum numbers are given in terms
of the initial variables it is useful to go back from $\zeta_\pm$
to the variables $r_a$. Using the relation \eqref{3s3} we find
\eq{
r_a^2=\frac{(\zeta_+-\omega_a^2)(\zeta_--\omega_a^2)}{\prod\limits_{b\neq{a}}^{\,}
(\omega_a^2-\omega_b^2)}=
\frac{(s_a^2-s_+^2)(s_a^2-s_-^2)}{\prod\limits_{b\neq{a}}^{\,}(\omega_a^2-\omega_b^2)}
=\frac{(s_a^2+s_+s_-)^2-s_a^2(s_++s_-)^2}{\prod\limits_{b\neq{a}}^{\,}(\omega_a^2-\omega_b^2)}.\label{3s26}
} One can easily check that the following relations hold
\begin{align}
&s_++s_-=\frac{s_2^2-s_3^2}{s_2A_2(\xi)-s_3A_3(\xi)}=\frac{\omega_2^2-\omega_3^2}{s_3A_3
(\xi)-s_2A_2(\xi)},
\label{3s27}\\
&s_+s_-=s_2s_3\frac{s_3A_2(\xi)-s_2A_3(\xi)}{s_3A_3(\xi)-s_2A_2(\xi)},\label{3s28}
\end{align}
 and this allows to recast $r_a$ into the form
\begin{align}
&r_1^2=\frac{\left[(\omega_1^2-\omega_2^2)s_3A_3(\xi)-(\omega_1^2-\omega_3^2)s_2A_2(\xi)
\right]^2+s_1^2(\omega_2^2-\omega_3^2)^2}
{(\omega_1^2-\omega_2^2)(\omega_1^2-\omega_3^2)(s_3A_3(\xi)-s_2A_2(\xi))^2},\label{3s29}\\
&r_2^2=\frac{(\omega_2^2-\omega_3^2)}{(\omega_1^2-\omega_2^2)}s_2^2
\frac{1-A_2^2(\xi)}{(s_3A_3(\xi)-s_2A_2(\xi))^2},\label{3s30}\\
&r_3^2=\frac{(\omega_2^2-\omega_3^2)}{(\omega_1^2-\omega_3^2)}s_3^2
\frac{A_3^2(\xi)-1}{(s_3A_3(\xi)-s_2A_2(\xi))^2}.\label{3s31}
\end{align}
We  note again that the choice of parameters
$\kappa=\omega_1,\quad\beta=-\frac{C_1}{\omega_1}$ in \eqref{3s14}
leads to a giant magnon solution, while choosing $\kappa=C_1,\quad
\beta=-\frac{\omega_1}{C_1}$ we find the single spike string
solution.

\paragraph{Energy and momenta of the three-spin spikes in $S^5$} \ \\

Let us turn now to the dispersion relations. The unified approach
of reduction to NR integrable system allows us to perform the
analysis in an universal way. Therefore, the calculation of the
dispersion relations closely follows  the 2-spin case.  There we
used the conformal constraints and \eqref{3s14} to fix the
parameters, namely to have single spike solution the parameters
were fixed to be
$\kappa=C_1,\,\,\beta=-\frac{\omega_1}{C_1},\,\,\beta^2>1$.

In the general case the energy and spins are defined as:
\begin{align}
&E=\frac{T\kappa}{\alpha}\int\limits^\infty_{-\infty}d\xi,
\label{3s32}\\
&J_a=\frac{T}{\alpha}\int\limits^\infty_{-\infty}\left[\beta\frac{C_a}{\alpha^2-\beta^2}+
\frac{\alpha^2\omega_a}{\alpha^2-\beta^2}r_a^2\right]\,d\xi,\quad
a=1,2,3.\label{3s33}
\end{align}
From the previous paragraph we learned that in order to have the
particular solution we are looking for one must choose the
parameters as in \eqref{3s14}. Therefore, from now on we take
$C_{2,3}=0,\,\,\alpha=1$ and the angular momenta $J_{2,3}$ are
simply
\begin{equation}
J_a=\frac{T}{1-\beta^2}\int\limits^\infty_{-\infty}\omega_ar_a^2(\xi)d\xi,\,\,\,\,a=2,3.
\label{3s34}
\end{equation}
The latter can be easily computed with the help of the explicit
expressions for $r_a(\xi)$, \eqref{3s30}, \eqref{3s31} and the
integrals
\begin{align}
&\int\limits^\infty_{-\infty}\frac{(1-\tanh^2{x})dx}{[\tanh{x}-c\coth(cx+b)]^2}=\frac{2}{c^2-1},
\label{3s35}\\
&\int\limits^\infty_{-\infty}\frac{(\coth^2(cx+b)-1)dx}{[\tanh{x}-c\coth(cx+b)]^2}=\frac{2}{c(c^2-1)}.
\label{3s36}
\end{align}
The result is:
\begin{equation}
\frac{1}{T}J_a=-\frac{2\omega_as_a}{\omega_1^2-\omega_a^2}
=-\frac{2\omega_a}{\omega_1^2-\omega_a^2}\sqrt{\bar{\zeta}-\omega_a^2},\,\,\,\,a=2,3.\label{3s37}
\end{equation}
On the other hand the expression for $J_1$ is more involved. Using
the values of the parameters fixed above we get:
\begin{equation}
J_1=\frac{TC_1^2}{\omega_1^2-C_1^2}\omega_1\int\limits^\infty_{-\infty}(1-r_1^2)d\xi.\label{3s38}
\end{equation}
Having in mind that $\sum\limits_{a=1}^3{r_a^2=1}$ and the
explicit form of $J_{2,3}$
\begin{align}
&J_2=-\frac{TC_1^2}{\omega_1^2-C_1^2}\omega_2\int\limits^\infty_{-\infty}r_2^2\,d\xi,\label{3s39}\\
&J_3=-\frac{TC_1^2}{\omega_1^2-C_1^2}\omega_3\int\limits^\infty_{-\infty}r_3^2\,d\xi,\label{3s40}
\end{align}
we find
\begin{equation}
J_1=\frac{TC_1^2}{\omega_1^2-C_1^2}\omega_1\int\limits^\infty_{-\infty}(r_2^2+r_3^2)d\xi
=-\omega_1\left[\frac{J_2}{\omega_2}+\frac{J_3}{\omega_3}\right].\label{3s41}
\end{equation}
As we pointed out in the beginning of this section, the
frequencies $\omega_a$ are ordered as follows:
$\omega_3<\omega_2<\omega_1$, which allows the following
definitions
\begin{equation}
\omega_a=\omega_1\sin\gamma_a,\qquad
0<\gamma_a<\frac{\pi}{2},\quad a=2,3.\label{3s42}
\end{equation}
In these notations we have the following relation between the
spins $J_a$ :
\begin{equation}
J_1=-\left(\frac{J_2}{\sin\gamma_2}+\frac{J_3}{\sin\gamma_3}\right).\label{3s43}
\end{equation}
The energy of the solution is simply
\begin{equation}
E=\frac{T\kappa}{\alpha}\int\limits^\infty_{-\infty}d\xi=TC_1\int\limits^\infty_{-\infty}d\xi.
\label{3s44}
\end{equation}
and it is obviously divergent. However the experience from the two
spin case teach us that one should look also at the angle $\phi_1$
around which the string is infinitely wound. The equation of
motion for $\phi_1(\xi)$ is given by
\begin{equation}
\phi'_1=\frac{1}{\alpha^2-\beta^2}\left[\frac{C_1}{r_1^2}+\beta\omega_1\right].\label{3s45}
\end{equation}
In addition we have the parameters fixed as $\kappa=C_1$ and
$\beta=-\frac{\omega_1}{C_1},\quad\beta^2>1,\quad\alpha=1$. Then
\begin{equation}
\Delta\phi_1=\frac{C_1}{C_1^2-\omega_1^2}\int\limits^\infty_{-\infty}
\frac{C_1^2-\omega_1^2r_1^2}{r_1^2}d\xi.\label{3s46}
\end{equation}
As in the two spin case, one can evaluate the following difference
expecting that it is finite
\begin{equation}
E-T\Delta\phi_1=-T\frac{C_1^3}{C_1^2-\omega_1^2}\int\limits^\infty_{-\infty}
\frac{1-r_1^2}{r_1^2}d\xi=-T\frac{C_1}{1-\beta^2}\int\limits^\infty_{-\infty}
\frac{1-r_1^2}{r_1^2}d\xi.\label{3s47}
\end{equation}
To prove this let us compute the difference  $E-T\phi_1$
explicitly  . First we mention the relations
\begin{align}
&\frac{d\zeta_+}{(\zeta_+-\omega_2^2)(\zeta_+-\omega_3^2)\sqrt{\bar{\zeta}-\zeta_+}}+
\frac{d\zeta_-}{(\zeta_--\omega_2^2)(\zeta_--\omega_3^2)\sqrt{\bar{\zeta}-\zeta_-}}=0,\label{3s48}\\
&\frac{\zeta_+\,d\zeta_+}{(\zeta_+-\omega_2^2)(\zeta_+-\omega_3^2)\sqrt{\bar{\zeta}-\zeta_+}}+
\frac{\zeta_-\,d\zeta_-}{(\zeta_--\omega_2^2)(\zeta_--\omega_3^2)\sqrt{\bar{\zeta}-\zeta_-}}=
-\frac{2\xi}{1-\beta^2},\label{3s49}
\end{align}
which allow to find the differential of the function
$[E-T\phi_1](\xi)$  in terms of the variables $\zeta_\pm$
\begin{equation}
d[E-T\phi_1](\xi)=-T\frac{C_1}{2}\left[-\frac{d\zeta_+}{(\zeta_+-\omega_1^2)\sqrt{\bar{\zeta}-\zeta_+}}-
\frac{d\zeta_-}{(\zeta_--\omega_1^2)\sqrt{\bar{\zeta}-\zeta_-}}\right].\label{3s50}
\end{equation}
Now we have simply to integrate over $\zeta_\pm$, the result is:
\begin{equation}
[E-T\Delta\phi_1](\xi)=T\left[\arctan\frac{\sqrt{\bar{\zeta}-\zeta_+}}{\sqrt{\omega_1^2-\bar{\zeta}}}+
\arctan\frac{\sqrt{\bar{\zeta}-\zeta_-}}{\sqrt{\omega_1^2-\bar{\zeta}}}\right].\label{3s51}
\end{equation}
Using the useful definitions \eqref{3s23} we find
\begin{equation}
[E-T\Delta\phi_1](\xi)=T\left[\arctan\frac{s_+}{s_1}+
\arctan\frac{s_-}{s_1}\right].\label{3s52}
\end{equation}
This can be written also as
\begin{equation}
[E-T\Delta\phi_1](\xi)=T\arctan\frac{s_1(s_++s_-)}{s_1^2-s_+s_-}.\label{3s53}
\end{equation}
Here we should note that as in the magnon case, this expression
was derived for a piece of the string. However as in the magnon
case it is straightforward be extended to all values
$-\infty<\xi<+\infty$. To do that it is important to carefully
take into account the asymptotic behavior of the quantities. In
particular, from \eqref{3s27}, \eqref{3s28}, \eqref{3s24} and
$\beta^2>1$ we learn that $(s_+s_-)(\pm\infty)=s_2s_3$ and
$(s_++s_-)(\pm\infty)=\pm(s_2+s_3)$. Therefore, we conclude that
\begin{multline}
E-T\Delta\phi_1=[E-T\phi_1](+\infty)-[E-T\phi_1](-\infty)=\\
=T\left[\arctan\left(\frac{s_1(s_2+s_3)}{s_1^2-s_2s_3}\right)-\arctan\left(-\frac{s_1(s_2+s_3)}
{s_1^2-s_2s_3}
\right)\right]=\\
=2T\arctan\left(\frac{s_1(s_2+s_3)}{s_1^2-s_2s_3}\right),\label{3s54}
\end{multline}
One can cast this expression in the form:
\begin{equation}
E-T\Delta\phi_1=2T\left[\arctan\frac{s_2}{s_1}+
\arctan\frac{s_3}{s_1}\right].\label{3s55}
\end{equation}
Note that when $C_1$ increase, the parameters
$\frac{s_{2,3}}{s_1}$ decrease, thus we  define two auxiliary
angles $\tilde\phi_{2,3}$ as follows
\begin{equation}
\cot\tilde\phi_a=\frac{s_a}{s_1},\qquad
0<\tilde\phi_a<\frac{\pi}{2},\qquad a=2,3.\label{3s56}
\end{equation}
In terms of these angles the finite expression $E-T\phi_1$ takes
the form
\begin{equation}
E-T\Delta\phi_1=2T\left[(\frac{\pi}{2}-\tilde\phi_2)+
(\frac{\pi}{2}-\tilde\phi_3)\right].\label{3s57}
\end{equation}
or, since $T=\frac{\sqrt{\lambda}}{2\pi}$,
\begin{equation}
E-T\Delta\phi_1=\frac{\sqrt{\lambda}}{\pi}\left[\bar{\theta}_2+
\bar{\theta}_3\right],\label{3s58}
\end{equation}
where  we defined $\bar{\theta}_a=\frac{\pi}{2}-\tilde\phi_a$.
Since
\begin{equation}
s_a=\sqrt{\omega_1^2-\omega_a^2}\sin\bar{\theta}_a,\label{3s59}
\end{equation}
and \eqref{3s42} for spins \eqref{3s37} we get
\begin{equation}
J_a=-2T\tan\gamma_a\sin\bar{\theta}_a.\label{3s60}
\end{equation}
Therefore from \eqref{3s43} we have
\begin{equation}
J_1=-\left(\frac{J_2}{\sin\gamma_2}+\frac{J_3}{\sin\gamma_3}\right)=
2T\left(\frac{\sin\bar{\theta}_2}{\cos\gamma_2}+\frac{\sin\bar{\theta}_3}{\cos\gamma_3}\right).
\label{3s61}
\end{equation}
Eliminating the angles $\gamma_a$ we obtain:
\begin{equation}
J_1=\sqrt{J_2^2+\frac{\lambda}{\pi^2}\sin\bar\theta_2}+\sqrt{J_3^2+\frac{\lambda}{\pi^2}\sin\bar\theta_3}.
\label{3s62}
\end{equation}
This result which shares some features with the case
of three spin giant magnons. The expressions are quite analogous
in sense of composition law. It was expected that this form will
hold in our case although this type of excitations are of infinite
energy but finite angular momenta! The finite expression \eqref{3s58}
is obtained extracting the infinite winding
number of the string configuration allowing only one spike. One
should note also that as in the single two spin cases, the angle
entering the dispersion relations is $\bar\theta_a=\pi/2-\theta_a$
instead of $\theta_a$ themselves. If we interpret the parameters
$\bar\theta_i$ as the half of quasimomenta $\frac{1}{2}p_i$ we will have complete analogy with
the magnon case interchanging the role of the enerfy and the momentum $J_1$. Indeed,
the definition of $\bar\theta_i$ is analogous to the definition of the correponding parameters.
in the magnon case. On this basis one can speculate on the relation to some spin chain, but
strong evidences in this direction are still missing.
Although it is not clear how
exactly and to which gauge theory operators these string solutions
are related, it will be interesting to further investigate this
type of solutions.

\subsection{Three-spin case: spins in both $AdS_5$ and $S^5$}

As we already discussed in the previous subsection, the singe
spike solutions are characterized with infinite energy
 and, in contrast with the magnon case, finite spins $J_a$. We studied so
far only the string solutions in the
spherical part $S^5$. It it is interesting to investigate the
multispin case where some of the spins are in the $AdS_5$ part. We
are going to address this question in the following subsection
concentrating on the case when we have two spin single spike
excitations in the spherical part and magnon-like excitation in
the $AdS_5$ part.

\paragraph{The ansatz} \ \\

The embedding of the string in the spherical part will be the same
as in the last subsection and for the $AdS_5$ part we
 will use the following ansatz
\al{
& Y_0=\m_0(\xi)e^{it}, \quad Y_1=\m_1(\xi)e^{i\Phi} \notag \\
& X_l=r_l(\xi)e^{i\tilde\phi_l}, \quad \xi=\a\sigma+\b\tau. } Here
we employ the following notations (we use $a,b=0,1$ and $l,m=1,2$)
\al{
& t=v_0\tau+\vf_0(\xi),  & \Phi=v_1\tau+\vf_1(\xi) & & \m_a=\m_a(\xi) \notag \\
& r_l=r_l(\xi), & \tilde\phi_1 =\omega_1\tau + \phi_1 (\xi) & &
\tilde\phi_2 =\omega_2\tau + \phi_2 (\xi)
\label{ansatz-1a}
}
In this parametrization the embedding is defined by the following
constraints (here $\kappa=v_0$)
\eq{ \e^{ab}\m_a\m_b=-1, \quad
r_1^2+r_2^2=1, \quad \e^{ab}=diag(-1,1),
}
where we explicitly assume restriction to $S^3\subset S^5$ and $AdS_3$. Having all
these definitions one can easily write the sigma model string
Lagrangian, which takes the form
\ml{
\L=(\a^2-\b^2)\e^{ab}\lb[\m'_a\m'_b+\m_a^2\lb(\vf_b'^2-\frac{v_b\b}{\a^2-\b^2}\rb)^2\rb]
- \frac{\a^2\e^{ab}}{(\a^2-\b^2)^2}v_a^2\m_a\m_b-\frac{\tilde\Lambda}{2}\lb(\e^{ab}\m_a\m_b+1\rb)\\
+(\a^2-\b^2)\d^{lm}\lb[r'_lr'_m+r_l^2\lb(\f_m'^2-\frac{\omega_m\b}{\a^2-\b^2}\rb)^2\rb]-
\frac{\a^2\d^{ab}}{(\a^2-\b^2)^2}\omega_a^2r_lr_m+\frac{\Lambda}{2}\lb(\d^{lm}r_lr_m-1\rb)
}
Following the general procedure of the reduction to the NR
system, we integrate once the equations of motion for $\vf_a$ and
$\f_l$ obtaining
\eq{
\vf_a'=\frac{1}{\a^2-\b^2}\lb[\frac{A_a}{\m_a^2}+v_a\b\rb], \quad
\f_a'=\frac{1}{\a^2-\b^2}\lb[\frac{C_l}{r_l^2}+\omega_l\b\rb]
\label{eq-phases}
}
The expressions for the conserved quantities can be easily read off:
\begin{align}
E\,=\,\frac{T}{\alpha}\int\limits^\infty_{-\infty}\mu_0^2(\xi)\left[\upsilon_0+\beta\varphi'_0(\xi)
\right]\,d\xi,\label{h1}\\
S\,=\,\frac{T}{\alpha}\int\limits^\infty_{-\infty}\mu_1^2(\xi)\left[\upsilon_1+\beta\varphi'_1(\xi)
\right]\,d\xi,\label{h2}\\
J_1\,=\,\frac{T}{\alpha}\int\limits^\infty_{-\infty}r_1^2(\xi)\left[\omega_1+\beta\phi'_1(\xi)\right]
\,d\xi,\label{h3}\\
J_2\,=\,\frac{T}{\alpha}\int\limits^\infty_{-\infty}r_2^2(\xi)\left[\omega_2+\beta\phi'_2(\xi)\right]
\,d\xi,\label{h4}
\end{align}
Using the equations of motion for $\varphi_a$ and $\phi_l$
(\ref{eq-phases}) and the constraints $\mu_1^2-\mu_0^2=-1,\,\quad
r_1^2+r_2^2=1$, these expressions can be written as
\begin{align}
&E\,=\,\frac{T}{\alpha}\,\frac{1}{(\alpha^2-\beta^2)}\int\limits^\infty_{-\infty}
\left[\alpha^2\upsilon_0\mu_0^2(\xi)+\beta\,A_0\right]\,d\xi,\label{h5}\\
&S\,=\,\frac{T}{\alpha}\,\frac{1}{(\alpha^2-\beta^2)}\int\limits^\infty_{-\infty}
\left[\alpha^2\upsilon_1(\mu_0^2-1)+\beta\,A_1\right]d\xi,\label{h6}\\
&J_1\,=\,\frac{T}{\alpha}\,\frac{1}{(\alpha^2-\beta^2)}\int\limits^\infty_{-\infty}
\left[\alpha^2\omega_1r_1^2(\xi)+\beta\,C_1\right]\,d\xi,\label{h7}\\
&J_2\,=\,\frac{T}{\alpha}\,\frac{1}{(\alpha^2-\beta^2)}\int\limits^\infty_{-\infty}
\left[\alpha^2\omega_2(1-r_1^2)+\beta\,C_2\right]d\xi.\label{h8}
\end{align}
We should also make use of the Virasoro constraints. In this case
these constraints are
\ml{
 (\a^2+\b^2)\e^{ab}\lb(\m_a'\m_b'+\m_a\m_b\vf_a'\vf_b'\rb) +\e^{ab}\m_a\m_b(v_b^2+2v_b\b\vf_b')\\
+(\a^2+\b^2)\d^{lm}\lb(r_l'r_m'+r_ar_b\f_l'\f_m'\rb)
+\d^{lm}r_lr_m(\omega_l^2+2\omega_l\b\f_m') =0 \label{imp-vir1} }
\eq{
\b\e^{ab}\lb(\m_a'\m_b'+\m_a\m_b\vf_a'\vf_b'\rb)+\e^{ab}v_a\m_a\m_b\vf_b'+
\b\d^{lm}\lb(r_l'r_m'+r_lr_m\f_l'\f_m'\rb)+\d^{lm}\omega_lr_lr_m\f_l'=0
\label{imp-vir2}
}
Using these expressions, the corresponding
equations of motion for $\m_a$ and $r_l$ can be written as
\al{
& (\a^2-\b^2)\m_a''-\frac{A_a}{(\a^2-\b^2)}\frac{1}{\m_a^3}
+\frac{\a^2\e^{ab}}{(\a^2-\b^2)} v_a^2\m_b+\tilde\Lambda\m_a=0
\label{eom-m}\\
& (\a^2-\b^2)r_l''-\frac{C_l}{(\a^2-\b^2)}\frac{1}{r_l^3}+\frac{\a^2\d^{lm}}{(\a^2-\b^2)}
\omega_l^2r_m+\Lambda r_l=0
\label{eom-r}
}
Again one can ask what is the effective Lagrangian leading to these equations of motion?
The answer is that these equations can be obtained from the Lagrangian
\ml{
\L=(\a^2-\b^2)\e^{ab}\m'_a\m'_b-\frac{\e^{ab}}{(\a^2-\b^2)}\frac{A_a^2}{\m_b^2}-
\frac{\a^2\e^{ab}}{(\a^2-\b^2)}v_a\m_b^2-\frac{\tilde\Lambda}{2}\lb(\e^{ab}\m_a\m_b+1\rb) \\
+(\a^2-\b^2)\d^{lm}r_l'r_m'-\frac{\d^{lm}}{(\a^2-\b^2)}\frac{C_l^2}{r_m^2}-
\frac{\a^2\d^{lm}}{(\a^2-\b^2)}\omega_lr_m^2-\frac{\Lambda}{2}\lb(\d^{lm}r_lr_m-1\rb)
}
which is, of course, NR integrable system.

Let us take closer look at the Virasoro constraints. One can
eliminate quadratic terms $\m_a'^2, r_l'^2, \vf_a'^2$ and
$\f_l'^2$ combining appropriately the two  Virasoro constraints
\ml{
 -\frac{(\a^2+\b^2)}{\b}\lb[\e^{ab}v_a\m_a\m_b\vf_a'+\d^{lm}\omega_lr_lr_m\f_m'\rb]\\
+\e^{ab}v_a^2\m_b^2 +\d^{lm}\omega_l^2r_m^2+2\e^{ab}\b
v_a\m_a\m_b\vf_b'+2\d^{lm}\omega_l\b r_lr_m\f_m'=0
}
or,
\eq{
\frac{(\a^2-\b^2)}{\b}\lb[\e^{ab}v_a\m_a\m_b\vf_a'+\d^{lm}\omega_lr_lr_m\f_m'\rb]=\e^{ab}v_a^2\m_b^2
+\d^{lm}\omega_l^2r_m^2
}
One can substitute in the last expression the solutions for $\vf_a'$ and $\f_l'$ to obtain  the
constraints
\eq{ \e^{ab}A_av_b+\d^{lm}C_l\omega_m=0
\label{constr-fin}
}
where $v_0=\kappa$. These are very important
and in order to have consistent dynamics should always be satisfied.

Let us return to the equations (\ref{eom-m}) and (\ref{eom-r}).
One can integrate them once getting
\al{
& \lb((\a^2-\b^2)\m_a'^2\rb)'+\lb(\frac{A_a^2}{(\a^2-\b^2)}\frac{1}{\m_a^2}\rb)'
+\lb(\frac{\a^2\e^{ab}}{(\a^2-\b^2)}v_a^2\m_b^2\rb)'+\tilde\Lambda\lb(\m_a^2\rb)'=0
\label{eom-m-1}\\
& \lb((\a^2-\b^2)r_l'^2\rb)'+\lb(\frac{C_l^2}{(\a^2-\b^2)}\frac{1}{r_l^2}\rb)'
+\lb(\frac{\a^2}{(\a^2-\b^2)}\omega_l^2r_l^2\rb)'+\Lambda\lb(r_l^2\rb)'=0
\label{eom-r-1}
}
 Since all the summations are with diagonal metrics
$\e^{ab}$ and $\d^{lm}$ one can sum correspondingly over the
indices the above equations to find
\al{
& \e^{ab}\lb[(\a^2-\b^2)\m_a'\m_b'+\frac{A_b^2}{(\a^2-\b^2)}\frac{1}{\m_a^2}
+\frac{\a^2}{(\a^2-\b^2)}v_a^2\m_b^2\rb]=H_a
\label{eom-m-2}\\
& \sum\limits_l\lb[(\a^2-\b^2)r_l'^2+\frac{C_l^2}{(\a^2-\b^2)}\frac{1}{r_l^2}
+\frac{\a^2}{(\a^2-\b^2)}\omega_l^2r_l^2\rb]=H_s
\label{eom-r-2}
}
where we used that $\e^{ab}\m_a\m_b=-1$ and $\d^{lm}r_lr_m=1$
(these eliminate the Lagrangian multipliers $\Lambda$ and
$\tilde\Lambda$ from the equations). We denoted the integration
constants by $H_a$ and $H_s$.

As we realized above, the constraint (\ref{constr-fin}) is not
that useful as in the case of motion in $S^5$ only. From the last
equations however we see that the spherical part and $AdS$ part
are actually separated. They look like  before but the constants
on the right hand side, $H_a$ and $H_s$, are related though the
Virasoro constraints. Roughly, the role of $\kappa$ (see for
instance (\ref{vir1-fin-1})), or of Hamiltonian in this case is
played by $H_a$ and $H_s$ correspondingly. This allows to use the
same method to obtain the solutions and dispersion relations, for
instance using (\ref{vir1-fin-nr}, \ref{vir2-fin-nr}) but now with
$H_a$ instead of $\kappa^2$.

Now let us return to the Virasoro constraints. Below we give very
detailed analysis of the constraints in this case.
 Multiplying
(\ref{imp-vir2}) by $2\b$ and subtracting it from (\ref{imp-vir1})
one finds
\ml{
(\a^2-\b^2)\e^{ab}\lb(\m_a'\m_b'+\m_a\m_b\vf_a'\vf_b'\rb) +\e^{ab}\m_a\m_bv_b^2\\
+(\a^2-\b^2)\d^{lm}\lb(r_l'r_m'+r_ar_b\f_l'\f_m'\rb)
+\d^{lm}r_lr_m\omega_l^2 =0
\label{imp-vir-h1}
}
Substituting for $\vf'$ and $\f'$ from (\ref{eq-phases}) we get
\ml{
(\a^2-\b^2)\e^{ab}\m_a'\m_b'+\e^{ab}\frac{A_a^2}{(\a^2-\b^2)}\frac{1}{\m_b^2}+
\frac{\a^2}{(\a^2-\b^2)}\e^{ab}v_a^2\m_b^2
+(\a^2-\b^2)\d^{lm}r_l'r_m'\\
+\d^{lm}\frac{C_l^2}{(\a^2-\b^2)}\frac{1}{r_m^2}+\frac{\a^2}{(\a^2-\b^2)}\d^{lm}\omega_l^2r_m^2
+\frac{2}{(\a^2-\b^2)}\left(\e^{ab}\b v_a
A_b+\d^{lm}\b\omega_lC_m\right)=0 }
which, using (\ref{eom-m-2},\ref{eom-r-2}), can be written as
\eq{
H_s+H_a+\frac{2}{(\a^2-\b^2)}\left(\e^{ab}\b v_a
A_b+\d^{lm}\b\omega_lC_m\right)=0
}
However due to (\ref{constr-fin}) the last two terms in the brackets vanish and
we end up with \eq{ H_a+H_s=0. \label{hamilton-0} } This is
actually the way the dynamics in the two part of the geometry is
related.

We now proceed with the analysis of $S^5$ part which is actually
reduced to $S^3\subset S^5$. Therefore we have to repeat the
analysis of two-spin case taking into account the above specifics.
To make considerations closer to that case we define\footnote{One
should not confuse $k$ with $\kappa$!}
\eq{
H_s=\frac{\a^2+\b^2}{\a^2-\b^2}k^2 .
\label{hamilton-k}
}
Choosing one of the turning points to be at $r_1=1$ (or $r_2=0$ which means
$\theta=\pi/2$). This means that in (\ref{eom-r-2}) one must
choose $C_2=0$. Indeed, one can use that $r_1^2+r_2^2=1$ and
express (\ref{eom-r-2}) in terms of $r_1$ only
\eq{
(\a^2-\b^2)\frac{r_1'^2}{1-r_1^2}+\frac{C_1^2}
{(\a^2-\b^2)}\frac{1}{r_1^2}+\frac{C_2^2}{(\a^2-\b^2)}
\frac{1}{1-r_1^2}
+\frac{\a^2}{(\a^2-\b^2)}\omega_1^2r_1^2+\frac{\a^2}{(\a^2-\b^2)}\omega_2^2(1-r_1^2)=H_s
}
which rules out non-trivial $C_2$. Next, expressing $\r'^2$ as
\eq{
(\a^2-\b^2)^2r_1'^2=\frac{1-r_1^2}{r_1^2}\lb[\{(\a^2+\b^2)k^2-\a^2\omega_2^2\}r_1^2-C_1^2-
\a^2(\omega_1^2-\omega_2^2)r_1^4\rb] .
\label{er1}
}
The second zero at $r_1=1$ can determine the relations between the parameters,
i.e. replacing $r_1=1$ in the square brackets on the right hand
side, we get
\eq{
(\a^2+\b^2)k^2=\a^2\omega_1^2+C_1^2
\label{C1}
}
One can take $\a^2=1$ as in a two-spin case, the spiky choice for
$k$ and other parameters is
\eq{
 k=C_1, \quad \beta=-\frac{\omega_1}{C_1}
.\label{beta}
}
 Then the equation \eqref{er1} can be rewrite as
\begin{equation}
\frac{(\alpha^2-\beta^2)^2}{4\alpha^2(\omega_1^2-\omega_2^2)}\left(\frac{dr_1^2}{d\xi}
\right)^2=(1-r_1^2)^2\left(r_1^2-\frac{C_1^2}{\alpha^2(\omega_1^2-\omega_2^2)}\right).\label{h9}
\end{equation}
Therefore
\begin{equation}
\frac{(\alpha^2-\beta^2)}{2\alpha\sqrt{\omega_1^2-\omega_2^2}}\,\frac{du}{d\xi}\,=-(1-u)\sqrt{u-\bar{u}},
\label{h10}
\end{equation}
or
\begin{equation}
d\xi\,=-\frac{(\alpha^2-\beta^2)}{2\alpha\sqrt{\omega_1^2-\omega_2^2}}\,\,\frac{du}{(1-u)\sqrt{u-\bar{u}}}\,,
\label{h11}
\end{equation}
where
\begin{equation}
u=r_1^2=\sin^2\theta,\qquad\qquad
\bar{u}=\frac{C_1^2}{\alpha^2(\omega_1^2-\omega_2^2)}=\sin^2\theta_0,\qquad\qquad\bar{u}<u<1.\label{h12}
\end{equation}
Now one can use the expression for the measure \eqref{h11} and the
parameter fixings $\alpha=1,\,\beta=-\frac{\omega_1}{C_1}$ to
calculate the conserved quantities $J_1$ and $J_2$ on the $S^5$
part. The result is
\begin{multline}
J_1\,=\,\frac{T\omega_1}{1-\beta^2}\,\int\limits^\infty_{-\infty}(u-1)\,d\xi=
\,\frac{2T\omega_1}{2\sqrt{\omega_1^2-\omega_2^2}}\,\int\limits_{\bar{u}}^{1}\frac{du}{\sqrt{u-\bar{u}}}=\\
=\,\frac{2T\omega_1}{\sqrt{\omega_1^2-\omega_2^2}}\sqrt{1-\bar{u}}=\frac{2T\omega_1}
{\sqrt{\omega_1^2-\omega_2^2}}\cos\theta_0\,=\\
=\,\frac{2T\omega_1}{\sqrt{\omega_1^2-\omega_2^2}}\sin\left(\frac{\pi}{2}-\theta_0\right),\\
J_2\,=\,-\,\frac{2T\omega_2}{\sqrt{\omega_1^2-\omega_2^2}}\cos\theta_0\,=\,-\,\frac{2T\omega_2}
{\sqrt{\omega_1^2-\omega_2^2}}\sin\left(\frac{\pi}{2}-\theta_0\right),\label{h13}
\end{multline}
or
\begin{equation}
\frac{J_1}{\omega_1}+\frac{J_2}{\omega_2}\,=\,0.\label{h14}
\end{equation}
Since $\omega_2<\omega_1$, one can define the ratio
$\omega_2/\omega_1$ as
$$\omega_2=\omega_1\sin\gamma\,.$$
In these notations the relation between the momenta $J_{1,2}$
becomes
\begin{equation}
J_1^2\,=\,J_2^2+\frac{\lambda}{\pi^2}\sin\left(\frac{\pi}{2}-\theta_0\right).\label{15}
\end{equation}
One can apply analogous analysis to the $AdS$ part of the
geometry. Let us write again the equation \eqref{eom-m-2} for
$\mu_0$
\ml{
(\a^2-\b^2)\frac{\m_0'^2}{\m_0^2-1}=\frac{1}{\m_0^2(\m_0^2-1)}\lb[H_a+\frac{A_0^2}{(\a^2-\b^2)\m_0^2}
+\frac{\a^2v_0^2}{(\a^2-\b^2)}\m_0^2 \rb. \\ \lb.
-\frac{A_1^2}{(\a^2-\b^2)(\m_0^2-1)}-\frac{\a^2v_1^2}{(\a^2-\b^2)}(\m_0^2-1)\rb],
}
which equivalently can be written as
\ml{
(\a^2-\b^2)\m_0'^2=\frac{1}{(\a^2-\b^2)\m_0^2}\lb[(\a^2-\b^2)\m_0^2(\m_0^2-1)H_a
\rb. \\ \lb.
+A_0^2(\m_0^2-1)+\a^2v_0^2\m_0^4(\m_0^2-1)-A_1^2\m_0^2-\a^2
v_1^2\m_0^2(\m_0^2-1)^2\rb] .
\label{eq-m0}
}
In the last subsection we discussed in some length the reasons for the choice
of the parameters. One can equally well repeat them here and find
that the appropriate choice for $A_1$ is $A_1=0$. It is then
useful to work in notations where $\m_0^2=z$ and analyze the pole
structure ensuring solutions of the type we are looking for.
Written in terms of $z$, the equation of motion contain
information for both parts - spherical and $AdS$, through the
conformal constraints. Below we give the details.

First of all it is useful to define $ \m_0^2=\cosh^2\rho=z $ and
rewrite the above equation \eqref{eq-m0} as
\begin{equation}
\frac{(\alpha^2-\beta^2)^2}{4}\left(\frac{dz}{d\xi}\right)^2=(z-1)
\left[(\alpha^2-\beta^2)H_az+A_0^2+\alpha^2v_0^2z^2-\alpha^2v_1^2(z-1)z\right].\label{h16}
\end{equation}
The requirement to have double zero in $z=1$ ensuring large energy
limit leads to the relation
\begin{equation}
(\alpha^2-\beta^2)k^2=A_0^2+\alpha^2v_0^2,\label{h17}
\end{equation}
 where we used the constraints \eqref{hamilton-0} and \eqref{hamilton-k}.
If we go back to the relation \eqref{C1}, one can easily find an
important relation between the parameters of $S^5$ and $AdS_5$
parts
\begin{equation}
\a^2\omega_1^2+C_1^2=A_0^2+\alpha^2v_0^2.\label{h18}
\end{equation}
Now, from the Virasoro constraints {\eqref{constr-fin}, one can
immediately find that $ v_0A_0=\omega_1 C_1$
 and with the help of the above relation one can to fix the parameters as follows
\begin{equation}
\alpha=1,\qquad v_0=C_1=k=1,\qquad A_0=\omega_1 \qquad\text{and
therefore}\qquad \beta=-\omega_1=-A_0.\label{h19}
\end{equation}
Then the equation \eqref{h16} can be rewritten in the form
\begin{equation}
\frac{(1-\beta^2)^2}{4(1-v_1^2)}\left(\frac{dz}{d\xi}\right)^2=
(z-1)^2\left(z-\frac{A_0^2}{1-v_1^2}\right),\label{h20}
\end{equation}
where we define $0<\bar{z}=\frac{A_0^2}{1-v_1^2}<1$. The
integration measure then is determined to be
\begin{equation}
d\xi=-\frac{(1-\beta^2)}{2\sqrt{1-v_1^2}}\frac{dz}{(z-1)\sqrt{z-\bar{z}}},\qquad
1<z<\infty.\label{h21}
\end{equation}
Using the expressions for the spin $S$ \eqref{h6}, the energy
\eqref{h5} and  \eqref{h19} we find the relations:
\begin{equation}
\frac{S}{v_1}\,=\,\frac{T}{(1-\beta^2)}\int\limits^\infty_{-\infty}(\mu_0^2-1)d\xi,\label{h22}
\end{equation}
\begin{multline}
E=\frac{T}{(1-\beta^2)}\int\limits^\infty_{-\infty}\left(\mu_0^2-A_0^2\right)d\xi=
\frac{T}{(1-\beta^2)}\int\limits^\infty_{-\infty}\left(\mu_0^2-1+1-A_0^2\right)d\xi=\\
=\,\frac{T}{(1-\beta^2)}\int\limits^\infty_{-\infty}\left(\mu_0^2-1\right)\,d\xi+
T\,\int\limits^\infty_{-\infty}\,d\xi.\label{h23}
\end{multline}
Combining all these we obtain the following relation between the
full energy and the spin on the $AdS$ part
\begin{equation}
E\,=\,\frac{S}{v_1}+T\,\int\limits^\infty_{-\infty}\,d\xi.\label{h24}
\end{equation}
As in the case of single spike in the spherical part, the
following difference is finite (see the two-spin case)
\begin{equation}
E-\frac{S}{v_1}-T\Delta\phi_1=T\int\limits^\infty_{-\infty}d\xi-T\Delta\phi_1,\label{h25}
\end{equation}
where $\Delta\phi_1=\int\limits^\infty_{-\infty}\phi'_1d\xi$ on
the $S^5$ part. To explicitly demonstrate the finiteness, let us
compute the difference
$T\int\limits^\infty_{-\infty}d\xi-T\Delta\phi_1$.\\
Using the equation for $\phi_1$
\begin{equation}
\phi'_1=\frac{1}{1-\beta^2}\left(\frac{C_1}{u}+\omega_1\beta\right),
\end{equation}
the expression for the integration measure $d\xi$ on the $S^5$
part \eqref{h11} and the relations \eqref{h19} we find
\begin{multline}
T\int\limits^\infty_{-\infty}d\xi-T\Delta\phi_1=\frac{T}{\sqrt{\omega_1^2-\omega_2^2}}
\int\limits_{\bar{u}}^{1}\left(\frac{1}{u}-1\right)\frac{du}{(u-1)\sqrt{u-\bar{u}}}=\\
=\frac{T}{\sqrt{\omega_1^2-\omega_2^2}}
\int\limits_{\bar{u}}^{1}\frac{du}{u\sqrt{u-\bar{u}}}=
\frac{2T}{\sqrt{\omega_1^2-\omega_2^2}}\frac{1}{\sqrt{\bar{u}}}\arctan\frac{\sqrt{1-\bar{u}}}{\sqrt{\bar{u}}}=\\
=2T\arctan\frac{\sqrt{1-\bar{u}}}{\sqrt{\bar{u}}}=2T\left(\frac{\pi}{2}-\theta_0\right)=2T\bar{\theta}.
\end{multline}
Obviously
\begin{equation}
E-T\Delta\phi_1=\frac{S}{v_1}+2T\bar{\theta}.\label{h26}
\end{equation}
On the other hand the $S$ spin contribution to the string energy
is expressed through \eqref{h21} as
\begin{equation}
\frac{S}{v_1}=\frac{T}{(1-\beta^2)}\int\limits^\infty_{-\infty}(\mu_0^2-1)d\xi=
-\frac{T}{\sqrt{1-v_1^2}}\int\limits_{1}^{\infty}\frac{dz}{\sqrt{z-\bar{z}}}.\label{h27}
\end{equation}
However, this integral diverges because the rotating long string
stretches to the boundary of $AdS_3$. Following the prescription
of ref. \cite{MTT} we subtract the divergent term to have a
regularized value $S_{\text{reg}}$
\begin{equation}
\frac{S_{\text{reg}}}{v_1}=\frac{2T}{\sqrt{1-v_1^2}}\sqrt{1-\bar{z}},\label{h28}
\end{equation}
from which $v_1$ is obtained to be
\begin{equation}
v_1=\frac{S_{\text{reg}}}{\sqrt{S_{\text{reg}}^2+4T^2(1-\bar{z})}}.\label{h29}
\end{equation}
Combining \eqref{h28} and \eqref{h29}, the $S$ spin contribution
is expressed as a magnon-like dispersion relation
\begin{equation}
\frac{S_{\text{reg}}}{v_1}=\sqrt{S_{\text{reg}}^2+4T^2(1-\bar{z})}.\label{h30}
\end{equation}
Here the parameter $0<\bar{z}=\frac{A_0^2}{1-v_1^2}<1$ is
characterized by a time difference between the two endpoints of
the string
\begin{equation}
{\Delta}t=\int\limits^\infty_{-\infty}\varphi'_0(\xi)d\xi=
2\arctan\frac{\sqrt{1-\bar{z}}}{\sqrt{\bar{z}}},
\end{equation}
which reduces to $\bar{z}=\cos^2\frac{{\Delta}t}{2}$.\\
Finally, we have the following dispersion relation
\begin{align}
E-T\Delta\phi_1& =
\sqrt{S_{\text{reg}}^2+\frac{\lambda}{\pi^2}\,\sin^2\frac{{\Delta}t}{2}}+\frac{\sqrt{\lambda}}
{\pi}\,\bar{\theta},\label{h31}\\
J_1^2&=J_2^2+\frac{\lambda}{\pi^2}\,\sin^2\bar{\theta}.\label{32}
\end{align}

Several comments are in order. First of all, we point out that the
string excitations in the two parts of the geometry are of
different type. While for those in the $AdS$ part there is good
understanding from the point of view of AdS/CFT, the excitation in
the spherical part of the geometry  are still not well understood.
One possible relation to the gauge theory suggests description in
terms of a Hubbard-like spin chain model, see for instance \cite{RSR}.

\sect{Conclusions}

In this paper we find and study multispin rotating string with
spikes. First we give concise review of ``spiky'' solutions of the
so called magnon type. These solutions are very important for
understanding the detailed mechanism of AdS/CFT correspondence.
They correspond to long gauge theory operators, as discussed in
the Introduction. The exact maps between dispersion relation of
string solutions and the anomalous dimensions of the gauge theory
operators is in the core of AdS/CFT correspondence and to
establish the latter it should be extended to wider energy scales.
Moreover, such studies  give strong arguments for conjecturing the
exact form of the scattering matrix governing the correspondence.
The latter is based on integrable structures on the both sides of
the correspondence. In view of this we approach the problem of
finding certain type of solutions, single spike solutions, using a
reduction to the Neumann-Rosochatius integrable system. It turns
out that all magnon type solutions can be obtained from this
system using a certain choice of the parameters describing the
families of solutions, which we demonstrated to be true for the
class of solutions we are interested in. A brief review of this
approach is presented in Section 2. In the next Section we give
some details of the two spin solutions of both types - magnon and
single spike solutions.

The main results of this paper are given in Section 4 where we
derive three spin spiky string solutions. There are two cases  we
investigate in this Section. First we obtain a three spin single
spike solution for strings extended only in the $S^5$ part of the
geometry. We find the dispersion relations in this case, which, on
the basis of AdS/CFT, is supposed to describe the anomalous
dimensions of certain gauge theory operators. The form of the
dispersion relation is similar ro that for the case of giant
magnons, but there is an essential difference. In this case all
the angular momenta are finite while the energy is still very
large. The finite quantity which make senses to deal with is the
``regularized'' expression for the energy, namely, we subtract
from the ``bare'' energy the (infinite) winding number given by
the angle difference $\Delta\phi$ and obtain a finite expression.
Physically this procedure makes sense, but since the angular
momenta are finite (and the energy infinite) one cannot directly
relate the results from the string side to certain gauge theory
operators. To do that one must study the upper part of the energy
spectrum which is conjectured to be described by Hubbard-like spin
chain \cite{Ishizeki:2007we,RSR,Roiban:2006jt,McLoughlin:2006cg,Beisert:06-nl}. 
Certainly it will be of great interest to
develop this approach further and find an appropriate mechanism to
describe this part of the spectrum. Another three spin solution we
investigate is a single spike string with two spins on $S^5$ and
one magnon excitation in the $AdS_5$ part of the space-time
geometry. Beside the issues described above, here we have an
additional magnon type excitation in $AdS_5$. The latter belongs
to the lower part of the spectrum but still has effective
(non-negligible) contribution to the dispersion relations. It
would be interesting to identify the gauge theory operators
corresponding to this type of string solutions.

Certainly the spiky string solutions are interesting not only because of
their close relation to the giant magnon solutions on the string
side. However their interpretation in the dual gauge theory is not
that obvious and needs further investigation.
 First of all, it would be interesting to further explore the power of integrable system
techniques and apply the "dressing method", used by Volovich and
Spradlin for giant magnons  (or B\"acklund transformations), to
find multi-solitonic extensions of the spiky strings. It could be
expected that the giant magnon and single spike solutions on
$AdS_3$ are related to the sinh-Gordon model. If so, it should be
relatively straightforward to study scattering in terms of the
latter and extract useful data which can be helpful for
predictions for the general case. The integrable structures
emerging in this identification can be indeed very useful. They
can be also used in certain sectors of the string/QCD
correspondence where the integrability is still preserved. Work on
some of these problems is in progress.

\bigskip
{\bf Note added:} After the submission of our paper to arXiv.org we learned about another paper on
strings with large winding \cite{Hayashi:2007bq}. The authors consider the two spin case
of single spike strings from the perspective of complex sine-Gordon model and give finite gap intepretations.

\bigskip
\leftline{\bf Acknowledgements}
\smallskip

We thank Nikolai Bobev for collaboration at the initial stage of this project,
for many discussions and critically reading the draft.
This work was supported in part by Bulgarian NSF BUF-14/06 grant.
 R.R. acknowledges  warm hospitality at the Institute for
Theoretical Physics, Vienna University of Technology where this project was mainly carried out. Many thanks
to Max Kreuzer and his group for fruitful and stimulating atmosphere. The work of R.R is partially supported
by Austrian Research Fund FWF grant \# P19051-N16.

\section{Appendix}
\appendix

For completeness and to make the paper self contained, here we
will give some details about the computations and techniques we
used in the paper.

\paragraph{About Neumann-Rosochatius system in brief} \ \\

Reduction to the $O(6)$ Neumann Rosochatius system can be found in \cite{arts-new-int}. One starts with the
string sigma model action
\eq{
\A=-\frac{\sqrt{\l}}{4\pi}\int d\sigma d\tau\, h^{\m\n} [G^{(AdS_5)}_{ab}\p_\m Y^a \p_\n Y^b
+G^{(S^5)}_{lm}\p_\m X^l \p_n X^m]
\label{app-action}
}
The following ansatz leads to the Neumann-Rosochatius integrable system
\al{
& Y^a=\hat\m^a (\sigma) e^{iv_a\tau}; \quad  Y^a\bar Y^b\eta_{ab}=-1\label{ans-nr-ads} \\
& X^l= \hat r^l (\sigma) e^{i\omega_l\tau}; \quad X^a\bar X^a=1 \label{ans-nr-s} \\
}
where the complex variables $\hat\m_a, \hat r_l$ are given by
\al{
& \hat \m_a=\m_a(\sigma) e^{\vf_a(\sigma)}, \quad \m_a\m_b\eta^{ab}=-1 \notag \\
& \hat r_l= r_l(\sigma) e^{\phi_l(\sigma)}, \quad r^a r^a=1 \label{ansatz-nr-phases}
}
For simplicity we will restrict ourself here to the case of $S^5$ since the $AdS$ part goes analogously.
The resulting Lagrangian coming from $S^5$ and defining Neuman-Rosochatius system  then is given by
\eq{
\L=\frac{1}{2}\sum\limits_{l=1}^3[{r'_l}^2-\omega^2 r^2_l-\frac{C_l^2}{r_l^2}]-\frac 12\Lambda
(\sum\limits_{l=1}^3r_l^2-1)
\label{nr-orig-ac}
}
In the above we used the equations of motion for the angles (integrated once)
\eq{
\phi'_l=\frac{C_l}{r_l^2}, \quad C_l \,\text{ are integration constants}
}
The Virasoro contraints can be formulated in terms of $r_l$ only, namely
\al{
& \kappa^2=\sum\limits_{l=1}^3{r'_l}^2+\omega^2 r^2_l+\frac{C_l^2}{r_l^2} \label{nr-vir-aa} \\
& \sum\limits_{l=1}^3\omega_l c_l =0 \label{nr-vir-bb}
}
The spins $J_l$ form the Cartan subalgebra of $SO(6)$ and are defined by
\eq{
J_l=\sqrt{\l}\,\omega_l\int\limits_0^{2\pi}\frac{d\sigma}{2\pi}\,r_l^2\equiv \sqrt{\l}\J_l
}
where the following relation holds
\eq{
\sum\limits_{l=1}^3\frac{\J_l}{\omega_l}=1
}
The classical integrals of motion are given by
\eq{
I_l=r^2_l+\sum\limits_{m\neq l}^3\frac{1}{\omega^2_l-\omega^2_m}
[(r_l r_m'-r_m r_l')^2 +\frac{C_l}{r^2_l}r_m^2 +
\frac{C_m}{r^2_m}r_l^2], \quad \sum\limits_{l=1}^3 I_l=1
\label{int-motion}
}
It is convenient to change the variables to such that respect the symmetry of the problem and then
separate the variables.
In our case the convenient coordinates $\zeta_\pm$ are the ellipsoidal ones defined by ($r_l^2=1$)
\eq{
r_l=\sqrt{\frac{(\omega_l^2-\zeta_-)(\omega_l^2-\zeta_+)}{\prod\limits_{m\neq l}^3(\omega_l^2-\omega_m^2)}}
}
Expressing the integrals of motion \eqref{int-motion} in terms of $\zeta_\pm$ we find the following
separable system of equations
\eq{
 \lb(\frac{d\zeta_-}{d\sigma}\rb)^2=-4\frac{P_5(\zeta_-)}{(\zeta_--\zeta_+)^2}
\quad \lb(\frac{d\zeta_+}{d\sigma}\rb)^2=-4\frac{P_5(\zeta_+)}{(\zeta_--\zeta_+)^2}
}
where $P_5(\zeta_\pm)$ is fifth degree polynomial containing the constants of motion. One can use
these equations to obtain the different types solutions we are looking for.

\paragraph{Parameter fixing in two spin cases} \ \\

Here we will present some details of the computations we skipped
in the main text, namely the choice of the parameters in the diffrent cases.

In general one can integrate once the equation of motion \eqref{eom-m}, \eqref{eom-m}
and analyze in which case one can find solutions of the type we are looking for.
To simplify the considerations we will restrict ourself to the case
of $S^3\subset S^5$. In this case we can even easily find the equation for, say, $r_1$,
the only independent dynamical variable, using the Virasoro constraints
(for instance \eqref{hamil-1}, \eqref{vir1-fin-nr} and \eqref{vir2-fin-nr}).
The Hamiltonian reads off
\eq{
H=(1-\b^2)\frac{{r_1'}^2}{1-r_1^2}+
\frac{1}{1-\b^2}\lb( \frac{C_1^2}{r_1^2}+ \frac{C_2^2}{1-r_1^2} \rb)
+\frac{\omega_1^2-\omega_2^2}{1-\b^2}r_1^2+\frac{\omega_2^2}{1-\b^2}
\label{hamil-app}
}
The equation of motion for $r_1$ is
\footnote{To simplify the notations we set $\a=1$, which does not affect the
considerations}
\ml{
(1-\b^2)^2{r_1'}^2=\frac{1}{r_1^2}[((1+\b^2)\kappa^2-
\omega_2^2)(1-r_1^2)r_1^2-C_1^2\\
+(C_1^2-C_2^2)r_1^2-(\omega_1^2-\omega_2^2)(1-r_1^2)r_1^4]
\label{eq-r-2sp-app}
}
The right hand side determines the turning points ${r_1'}^2=0$. We observe
that they are three and one should determine them from physical conditions.
Since we are in the lower region of the energy scale the excitations are near
the null geodesic, i.e. the equator. Therefore, we want our string
to be extended to the equator of the sphere, $r_1=1$. From the explicit form of the Hamiltonian
\eqref{hamil-app} one can expect that $C_2$ must be zero. Indeed, requiring
that one of the turning points is $r_1=1$ one can find that it may happen only if $C_2=0$.
The equation \eqref{eq-r-2sp-app} then greatly simplifies,
\eq{
(1-\b^2)^2{r_1'}^2=\frac{(1-r_1^2)}{r_1^2}
[((1+\b^2)\kappa^2--\omega_2^2)r_1^2-C_1^2-(\omega_1^2-\omega_2^2)r_1^4]
\label{eq-e-nr-app1}
}
 but still we have two more zeroes on the right hand side. Therefore we need $r_1=1$ to
be double zero. The latter condition leads to the constraints
\eq{
(1+\b^2)\kappa^2=\omega_1^2+C_1^2, \quad \omega_1 C_1+\omega_2 C_2
+\b\kappa^2=0
\label{eq-param-nr-app}
}
With this fixing \eqref{eq-e-nr-app1} becomes
\eq{
(1-\b^2)^2{r_1'}^2=\frac{(1-r_1^2)^2}{r_1^2}(\omega_1^2-\omega_2^2)(r_1^2-\bar r_1^2)
\label{eq-e-nr-app2},
}
where the second turning point $\bar r_1$ is given by
\eq{
\bar r_1=\frac{C_1}{\sqrt{\omega_1^2-\omega_2^2}}.
}
Let us consider the constraints in more details. Having $C_2=0$ fixed, one can find for the parameter
$\b=-\frac{\omega_1 C_1}{\kappa^2}$. On other hand  one can solve the constraints \eqref{eq-param-nr-app}
for $\kappa$, but it has two solutions
\eq{
\kappa=
\begin{cases}
\omega_1\quad {\text{corresponding to a magnon solution}} \\
C_1 \quad {\text{corresponding to a single spike solution}}
\end{cases}
}

Let us give some details for the case of a two spin single spike solution.
First of all let us mention the equation for $r_1$ we get with the above
choice of the parameters
\eq{
(1-\b^2)^2{r_1'}^2=\frac{(1-r_1^2)^2}{r_1^2}\Delta\omega^2(r_1^2-\bar
r_1^2) \label{star1a}
}
(we mention that $\kappa=C_1, \,\, \b=-\frac{\omega_1}{C_1}, \,\, \Delta\omega^2=\omega_1^2-\omega_2^2$).

Defining
\eq{ u:=r_1^2=\sin^2\theta, \quad \sqrt{\bar
u}=\frac{C_1}{\sqrt{\Delta\omega}}
}
we get
\eq{
u'=-\frac{2\sqrt{\Delta\omega}(1-u)\sqrt{u-\bar u}}{1-\b^2}
}
Then the measure becomes
\eq{
dy=-\frac{(1-\b^2)du}{2\sqrt{\Delta\omega}(1-u)\sqrt{u-\bar u}},
\quad \bar u\leq u\leq 1
}
Some comments are in order. We choose the minus sign taking square root of
(\ref{star1}) because $\b\geq \a=1$ and starting from $\bar u$ the derivative
increases in order realizing a spike pointing towards the north
pole. Actually $u$ runs from $\bar u$ to 1 and back to $\bar u$ realizing
in the meantime (infinte) windings around $\phi_1$.

To obtain the angle difference  $\Delta\phi_1$ (actually this is the winding number) we have
to evaluate the integral
\al{ &
T\Delta\phi_1=\frac{C_1^3T}{C_1^2-\omega_1^2}\int\frac{dy}{u}-\frac{\omega_1^2C_1}
{C_1^2-\omega_1^2}\int dy \notag \\
& = \frac{C_1^3T}{C_1^2-\omega_1^2}\int\frac{1-u+u}{u}dy-\frac{\omega_1^2}{C_1^2-\omega_1^2}\ep_s \notag \\
&=\frac{C_1^3T}{C_1^2-\omega_1^2}\int\frac{1-u}{u}dy +E,
}
where we defined
\eq{
E =TC_1\int dy
}
Now we are going to show that the following difference is finite
\eq{
E-T\Delta\phi_1=-\frac{C_1^3T}{C_1^2-\omega_1^2}\int\frac{1-u}{u}dy.
}
Passing to $u$ variable, we perform the integration to find the
winding $\Delta\Phi_1$, i.e
\eq{
E-T\Delta\phi_1=\frac{C_1T}{\sqrt{\Delta\omega^2}}\int
\frac{du}{u\sqrt{u-\bar u}}=2T\,\arccos\sqrt{\bar u}
}
Since $\bar u=\sin^2\theta_0$ the solution can be written as
\eq{
E-T\Delta\phi_1=\frac{\sqrt{\l}}{\pi}(\frac{\pi}{2}-\theta_0)=
\frac{\sqrt{\l}}{\pi}\bar\theta_0,
}
which is the final answer.

One can extend the considerations to the full $S^5$ part or $SdS_5$ part of the geometry
using the same technology.

\end{document}